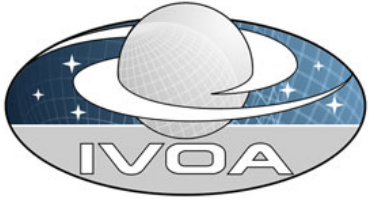

# VOSpace specification
# Version 2.00-20133029

## IVOA Recommendation 2013 March 29

**This version:**
http://www.ivoa.net/Documents/WD/GWS/REC-VOSpace-2.0-20130329.html
**Latest version:**
http://www.ivoa.net/Documents/latest/VOSpace-2.0.html
**Previous versions:**
http://www.ivoa.net/Documents/WD/GWS/PR-VOSpace-2.0-20121221.html
http://www.ivoa.net/Documents/WD/GWS/PR-VOSpace-2.0-20120824.html
PR: http://www.ivoa.net/Documents/WD/GWS/PR-VOSpace-2.0-20111202.html
WD: http://www.ivoa.net/Documents/WD/GWS/WD-VOSpace-2.0-20110628.html
WD: http://www.ivoa.net/Documents/WD/GWS/WD-VOSpace-2.0-20101112.html
WD: http://www.ivoa.net/Documents/GWS/WD-VOSpace-2.0-20100323.html
WD: http://www.ivoa.net/Documents/GWS/WD-VOSpace-2.0-20090904.html

WD: http://www.ivoa.net/Documents/GWS/WD-VOSpace-2.0-20090513.doc

**Working Group:**
http://www.ivoa.net/twiki/bin/view/IVOA/IvoaGridAndWebServices
**Author(s):**
Matthew Graham (editor)
Dave Morris
Guy Rixon
Pat Dowler
Andre Schaaff
Doug Tody

## Abstract

VOSpace is the IVOA interface to distributed storage. This specification presents the first RESTful version of the interface, which is functionally equivalent to the SOAP-based VOSpace 1.1 specification. Note that all prior VOSpace clients will not work with this new version of the interface.

## Status of this Document

This document has been produced by the IVOA Grid and Web Services Working Group.
It has been reviewed by IVOA Members and other interested parties, and has been endorsed by the IVOA Executive Committee as an IVOA Recommendation. It is a stable document and may be used as reference material or cited as a normative reference from another document. IVOA's role in making the Recommendation is to draw attention to the specification and to promote its widespread deployment. This enhances the functionality and interoperability inside the Astronomical Community.

A list of current IVOA Recommendations and other technical documents can be found at http://www.ivoa.net/Documents/.

## Acknowledgments

This document derives from discussions among the Grid and Web Services working group of the IVOA.

This document has been developed with support from the National Science Foundation's Information Technology Research Program under Cooperative Agreement AST0122449 with the John Hopkins University, from the UK Science and Technology Facilities Council (STFC), and from the European Commission's Sixth Framework Program via the Optical Infrared Coordination Network (OPTICON).

## Conformance related definitions

The words "MUST", "SHALL", "SHOULD", "MAY", "RECOMMENDED", and "OPTIONAL" (in upper or lower case) used in this document are to be interpreted as described in IETF standard, RFC 2119 [RFC 2119].

The **Virtual Observatory (VO)** is a general term for a collection of federated resources that can be used to conduct astronomical research, education, and outreach. The **International Virtual Observatory Alliance (IVOA)** is a global collaboration of separately funded projects to develop standards and infrastructure that enable VO applications. The International Virtual Observatory (IVO) application is an application that takes advantage of IVOA standards and infrastructure to provide some VO service.

## Contents

## 1. Introduction

VOSpace is the IVOA interface to distributed storage. It specifies how VO agents and applications can use network attached data stores to persist and exchange data in a standard way.

A VOSpace web service is an access point for a distributed storage network. Through this access point, a client can:

- add or delete data objects
- manipulate metadata for the data objects
- obtain URIs through which the content of the data objects can be accessed

VOSpace does not define how the data is stored or transferred, only the control messages to gain access. Thus, the VOSpace interface can readily be added to an existing storage system.

When we speak of "a VOSpace", we mean the arrangement of data accessible through one particular VOSpace service.

Each data object within a VOSpace service is represented as a node and has a description called a representation. A useful analogy to have in mind when reading this document is that a node is equivalent to a file.

Nodes in VOSpace have unique identifiers expressed as URIs in the 'vos' scheme, as defined below.

VOSpace 2.0 does not introduce any new functionality to that already offered by prior (SOAP-based) versions of the interface (VOSpace 1.1) but defines a RESTful binding for the interface.

## 1.1 Typical use of a VOSpace service

A typical use case for VOSpace is uploading a local data file to a remote VOSpace service. This is a two-stage process: creating a description of the data file (representation) in the VOSpace including any metadata (its properties) that they want to associate with it (e.g., MIME type), and defining the transfer operation that will actually see the data file bytes uploaded to the VOSpace service. The order of the processes should not matter. The user may want to create the representation first and then perform the transfer or transfer the bytes first and then update the representation with the appropriate metadata.

Let's consider the first sequence: the user provides a XML description of the data file which they HTTP PUT to the appropriate VOSpace URI - this will be the HTTP identifier for the data file in the VOSpace, e.g. http://nvo.caltech.edu/vospace/myData/table123. The description will resemble this:

```
<node xmlns="http://www.ivoa.net/xml/VOSpaceTypes-v2.0"
      xmlns:xsi="http://www.w3.org/2001/XMLSchema-instance"
      uri="vos://nvo.caltech!vospace/mytable1"
      xsi:type="vost:UnstructuredDataNode">
    <properties>
        <property uri="ivo://ivoa.net/vospace/core#mimetype">text/xml</property>
    </properties>
</node>
```

The service will reply with an amended version of the representation containing service-specific details in addition to the information supplied by the user. These will include data formats that the service can handle for the type of node created in the VOSpace, third-party interfaces (capabilities) to the data that the service offers and system metadata.

The user will then describe the data format (the view) they want to use in uploading the file, e.g. VOTable, and the transport protocol (the protocol) that they want to employ to upload the file, e.g. HTTP PUT. This will result in the HTTP POSTing of a XML description of the transfer request to the appropriate VOSpace URI, e.g. http://nvo.caltech.edu/vospace/myData/table123/transfers. The description will resemble this:

```
<transfer xmlns="http://www.ivoa.net/xml/VOSpaceTypes-v2.0">
    <direction>pushToVoSpace</direction>
    <view uri="ivo://ivoa.net/vospace/core#votable" />
    <protocol uri="ivo://ivoa.net/vospace/core#http-put" />
</transfer>
```

The service will reply with the URL that the user will HTTP PUT the data file to, e.g. http://nvo.caltech.edu/bvospace/myData/table123/transfers/147516ab. The user will then use a regular HTTP client to transfer (PUT) the local file to the specified endpoint. This illustrates an important point about VOSpace - it is only concerned with the server-side management of data storage and transfer. A client negotiates the details of a data transfer with a VOSpace service but the actual transfer of bytes across a network is handled by other tools.

Similarly, when a user wants to retrieve a data file from a VOSpace service, they will specify the data format (view) they want to use in downloading the file, e.g. VOTable, and the transport protocol (the protocol) that they want to employ to download the file, e.g. HTTP GET, and HTTP POST a XML description of this transfer request to the appropriate VOSpace URI - the transfer URI for the node in the VOSpace, e.g. http://nvo.caltech.edu/vospace/myDataNode/table123/transfers. The description will resemble this:

```
<transfer xmlns="http://www.ivoa.net/xml/VOSpaceTypes-v2.0">
    <direction>pullFromVoSpace</direction>
    <view uri="ivo://ivoa.net/vospace/core#votable" />
    <protocol uri="ivo://ivoa.net/vospace/core#httpget" />
</transfer>
```

The service will reply with the URL for the user to use, e.g. http://nvo.caltech.edu/vospace/myDataNode/table123/transfers/3df89ab4. The user can then download the data file by pointing an HTTP client (e.g. web browser) at the specified endpoint.

## 1.2 The role in the IVOA Architecture

The IVOA Architecture [Arch] provides a high-level view of how IVOA standards work together to connect users and applications with providers of data and services, as depicted in the diagram in Fig. 1.

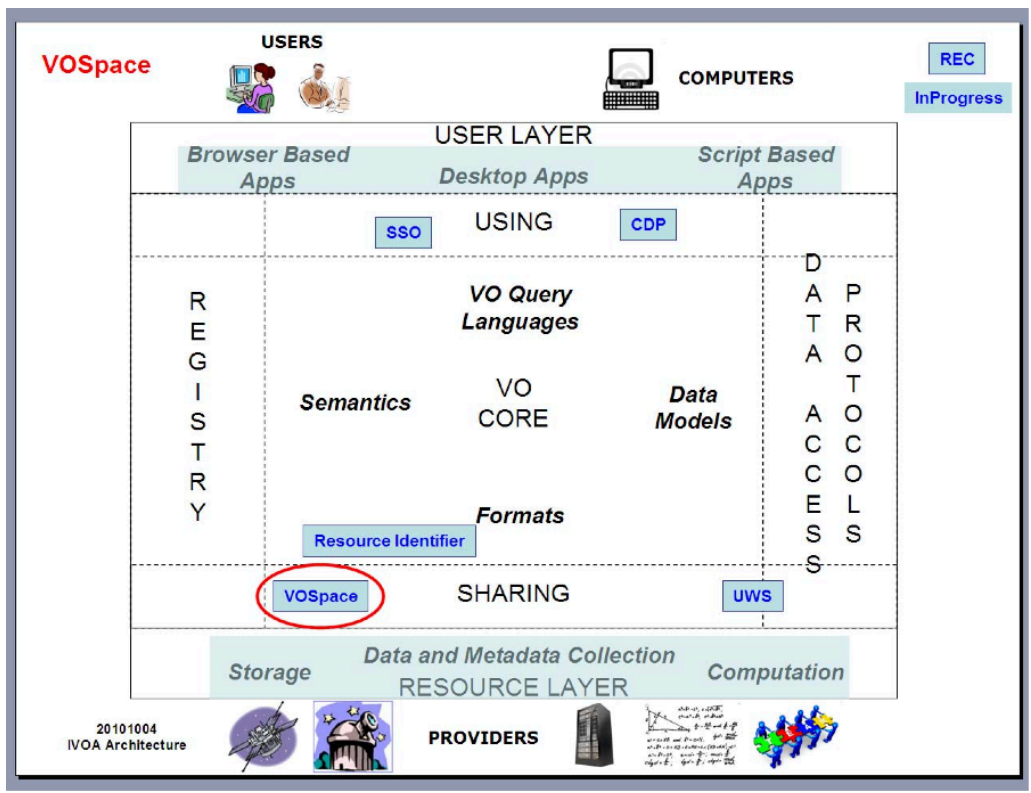

**Figure 1. VOSpace in the IVOA Architecture.** This provides an interface to distributed storage. It specifies how applications can use networked data stores to persist and exchange data in a standardized fashion.

In this architecture, users employ a variety of tools (from the User Layer) to discover and access archives and services of interest (represented in the Resource Layer). VOSpace provides an interface to storage resources containing the results of using these archives and services and also to other storage solutions, e.g., local disks, where users might want to transfer these results for further work. Items in these resources are referenced by a VOSpace identifier which is related to the standard IVOA Resource Identifier (see section 2). This version of VOSpace employs the UWS design pattern [UWS] to manage data transfers (see section 3.6) and searches (see section 3.7). VOSpace instances may also employ the IVOA Single-Sign-On standard [SSO] for authentication purposes (see section 4) and IVOA Credential Delegation Protocol [CDP] to delegate data transfers.

## 1.3 Document roadmap

The rest of this document is structured as follows:

In Section 2, we specify the URI syntax for identifying data objects (nodes) in VOSpace.

In Section 3, we present the data model that underpins the VOSpace architecture. This consists of a number of data structures, which have XML representations that are used across the wire in message exchanges with a VOSpace service. These structures represent:

- the data objects themselves (nodes)
- metadata that can be associated with a data object (properties)
- third-party interfaces to the data (capabilities)
- the data format used when transferring data objects across the wire (views)
- the transport protocol employed in a data transfer (protocols)
- the data transfer itself (transfers)
- searches of data objects (searches)

We also describe the REST bindings between these representations and their URIs (HTTP identifiers).

In Section 4, we outline how security and access control policies are currently handled in VOSpace.

In Section 5, we detail the operations that the VOSpace interface supports. These handle access to service-level metadata, the creation and manipulation of nodes within the VOSpace, access to node metadata (properties) and data transfer to and from the VOSpace.

In Appendix A, we formally define the VOSpace interface with a machine readable description of its requests and responses and in Appendix B, we present a compliance matrix listing the mandatory behaviour required of a valid VOSpace 2.0 service.

## 2. VOSpace identifiers

The identifier for a node in VOSpace SHALL be a URI with the scheme `vos`.

Such a URI SHALL have the following parts with the meanings and encoding rules defined in RFC2396 [RFC 2396].

- scheme

- naming authority
- path
- (optional) query
- (optional) fragment identifier (with the expected semantics [see here])

The naming authority for a VOSpace node SHALL be the VOSpace service through which the node was created. The authority part of the URI SHALL be constructed from the IVO registry identifier [IVORN] for that service by deleting the ivo:// prefix and changing all forward-slash characters('/') in the resource key to exclamation marks ('!') or tildes ('~'). Note that a service SHALL be consistent in its use of separator characters ('!' or '~') when referring to its own data but SHALL accept either as valid in URIs in service requests. For the rest of the document, we shall use '!' as the default character.

This is an example of a possible VOSpace identifier.

```
vos://nvo.caltech!vospace/myresults/siap-out-1.vot
```

The URI scheme is *vos*

Using a separate URI scheme for VOSpace identifiers enables clients to distinguish between IVO registry identifiers and VOSpace identifiers.

- nvo.caltech!vospace

is the authority part of the URI, corresponding to the IVO registry identifier

- ivo://nvo.caltech/vospace

This is the IVO registry identifier of the VOSpace service that contains the node.

- /siap-out-1.vot is the URI path

Slashes in the URI path imply a hierarchical arrangement of data: the data object identified by vos://nvo.caltech!vospace/myresults/siap-out-1.vot is within the container identified by vos://nvo.caltech!vospace/myresults.

Literal space characters are also not allowed in URIs.

All ancestors in the hierarchy SHALL be resolvable as containers (ContainerNodes), all the way up to the root node of the space (this precludes any system of implied hierarchy in the naming scheme for nodes with ancestors that are just logical entities and cannot be reified, e.g. the Amazon S3 system).

A VOSpace identifier is globally unique, and identifies one specific node in a specific VOSpace service.

The standardID for this specification SHALL be: ivo://ivoa.net/std/VOSpace/v2.0.

## 2.1 Identifier resolution

A VOSpace identifier can be resolved to a HTTP endpoint for accessing representations of the node associated with it. A client SHOULD use the following procedure to resolve access to a VOSpace node from a VOSpace identifier:

- Resolve HTTP service endpoint of VOSpace service with registry
- Append "nodes/" and the path following the naming authority part of the VOSpace identifier to the service endpoint

Given the example identifier

```
vos://org.astrogrid.cam!vospace/container-6/siap-out-1.vot?foo=bar
```

processing the URI to resolve the VOSpace service would involve :

- Extract the IVO registry identifier of the VOSpace service by prepending an ivo scheme to the naming authority part:
  - ivo://org.astrogrid.cam/vospace
- Resolve the IVO identifier in a registry and retrieve the access URL of the service endpoint:
  - http://some.uni.ac.uk/vospace
- Append "nodes/" and the path part of the VOSpace identifier:
  - http://some.uni.ac.uk/vospace/nodes/container-6/siap-out-1.vot?foo=bar

Note that any fragment identifier in the identifier SHOULD be removed when resolving the identifier to a HTTP endpoint, consistent with the implied semantics of URI fragments [see here].

# 3. VOSpace data model

## 3.1 Nodes and node types

We refer to the arrangement of data accessible through one particular VOSpace service as "a VOSpace".

Each data object within a VOSpace SHALL be represented as a node that is identified by a URI.

There are different types of nodes and the type of a VOSpace node determines how the VOSpace service stores and interprets the node data.

The types are arranged in a hierarchy (see Fig. 2), with more detailed types inheriting the structure of more generic types.

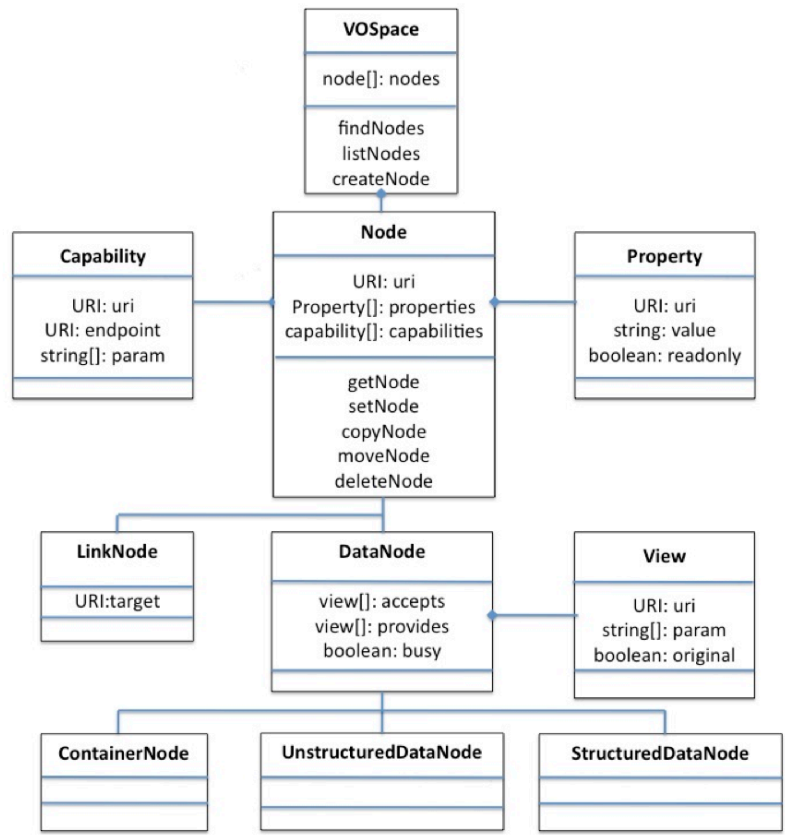

**Figure 2. Node hierarchy** This shows the inheritance structure for the different types of nodes in VOSpace.

The following types (and representations) are defined:

- *Node* is the most basic type
- *ContainerNode* describes a data item that can contain other data items
- *DataNode* describes a data item stored in the VOSpace
- *UnstructuredDataNode* describes a data item for which the VOSpace does not understand the data format
- *StructuredDataNode* describes a data item for which the space understands the format and may make transformations that preserve the meaning of the data.
- *LinkNode* describes a node that points to another node.

When data is stored and retrieved from an *UnstructuredDataNode*, the bit pattern read back SHALL be identical to that written.

When data is stored and retrieved from a *StructuredDataNode*, the bit pattern returned MAY be different to the original. For example, storing tabular data from a VOTable file will preserve the tabular data, but any comments in the original XML file may be lost.

A Node representation SHALL have the following elements:

- *uri*: the vos:// identifier for the node, URI-encoded according to RFC2396
- *properties*: a set of metadata properties for the node
- *capabilities*: a third-party interface to a data object

In addition, a *DataNode* representation SHALL have the following elements:

- *accepts*: a list of the views (data formats) that the node can accept
- *provides*: a list of the views (data formats) that the node can provide
- *busy*: a boolean flag to indicate that the data associated with the node cannot be accessed

The *busy* flag is used to indicate that an internal operation is in progress, and the node data is not available.

A *ContainerNode* representation SHALL have the following elements, in addition to those it inherits from the *Node* representation:

- *nodes*: a list of the direct children, if applicable, that the container has. Each child is represented as a *node* subelement containing its vos:// identifier, URI-encoded according to RFC2396

A *LinkNode* representation SHALL have the following elements, in addition to those it inherits from the *Node* representation:

- *target*: the target URI, URI-encoded according to RFC2396

The link target can be a URI that points to any type of resource, including other VOSpace Nodes (within the same VOSpace service or in another service), or external resources outside VOSpace altogether.

The properties of a *LinkNode* do not propagate to the target of the *LinkNode*, i.e., a property attached to a *LinkNode* does not also get attached to the target node. One use case is to enable third-party annotations to be associated with a resource but without the resource itself getting cluttered with unnecessary metadata. In this case, the client creates a *LinkNode* pointing to the resource in question and then adds the annotations as properties of the LinkNode.

Both the *ContainerNode* and the *LinkNode* SHALL have no data bytes associated with them.

The set of node types defined by this standard is closed; new types may be introduced only via new versions of the standard.

To comply with the standard, a client or service SHALL be able to parse XML representations of all the node types defined in the current specification.

Note: This does not require all services to support all of the Node types, just that it can process an XML request containing any of the types. If the service receives a request for a type that it does not support, the service SHOULD return a *TypeNotSupported* fault. The service SHALL NOT throw an XML parser error if it receives a request for a type that it does not support.

## 3.2 Properties

*Properties* are simple string-based metadata properties associated with a node.

Individual *Properties* should contain simple short string values, not large blocks of information. If a system needs to attach a large amount of metadata to a node, then it should either use multiple small *Properties*, or a single *Property* containing a URI or URL pointing to an external resource that contains the additional metadata.

A *Property* representation SHALL have the following elements:

- *uri*: the *Property* identifier
- *value*: the string value of the *Property*
- *readOnly*: a boolean flag to indicate that the *Property* cannot be changed by the client

Properties may be set by the client or the service.

### 3.2.1 Property values

Unless they have special meaning to the service or client, *Properties* are treated as simple string values.

When a *Property* can take multiple values, e.g., a list of groups which can access a particular resource, these SHOULD be comma-separated, unless the property description defines a specific delimiter.

Some *Properties* may have meaning to the service; others may have meaning only to one specific type of client. A service implementation does not need to understand the meaning of all the *Properties* of a node. Any *Properties* that it does not understand can simply be stored as text strings.

### 3.2.2 Property identifiers

Every new type of *Property* SHALL require a unique URI to identify the *Property* and its meaning.

The rules for the *Property* identifiers are similar to the rules for namespace URIs in XML schema. The only restriction is that it SHALL be a valid (unique) URI.

- An XML schema namespace identifier can be just a simple URN, e.g. urn:mynamespace
- Within the IVOA, the convention for namespace identifiers is to use a HTTP URL pointing to the namespace schema or a resource describing it

The current VOSpace schema defines *Property* identifiers as anyURI. The only restriction is that it SHALL be a valid (unique) URI.

- A *Property* URI can be a simple URN, e.g. urn:myproperty

This may be sufficient for testing and development on a private system, but it is not scalable for use on a public service.

For a production system, any new Properties SHOULD have unique URIs that can be resolved into a description of the Property.

Ideally, these should be IVO registry URIs that point to a description registered in the IVO registry:

- ivo://myregistry/vospace/properties#myproperty

Using an IVO registry URI to identify Properties has two main advantages :

- IVO registry URIs are by their nature unique, which makes it easy to ensure that different teams do not accidentally use the same URI
- If the IVO registry URI points to a description registered in the IVO registry, this provides a mechanism to discover what the Property means

### 3.2.3 Property descriptions

If the URI for a particular Property is resolvable, i.e. an IVO registry identifier or a HTTP URL, then it SHOULD point to an XML resource that describes the Property.

A Property description SHOULD describe the data type and meaning of a Property.

A PropertyDescription SHOULD have the following members :

- *uri*: the formal URI of the Property
- *DisplayName*: A display name for the Property
- *Description*: A text block describing the meaning and validation rules of the *Property*

A *PropertyDescription* MAY have the following OPTIONAL members :

- *UCD*: the Universal Content Descriptor (in the UCD1+ scheme) for the *Property*
- *Unit*: the unit of measurement of the Property

The information in a Property description can be used to generate a UI for displaying and modifying the different types of Properties.

Note that at the time of writing, the schema for registering PropertyDescriptions in the IVO registry has not been finalized.

### 3.2.3.1 UI display name

If a client is unable to resolve a Property identifier into a description, then it may just display the identifier as a text string:

- urn:modifieddate

If the client can resolve the Property identifier into a description, then the client may use the information in the description to display a human readable name and description of the Property:

- Last modification date of the node data

### 3.2.3.2 Property editors

If the client is unable to resolve a Property identifier into a description, or does not understand the type information defined in the description, then the client MAY treat the Property value as a simple text string.

If the client can resolve the Property identifier into a description, then the client MAY use the information in the description to display an appropriate editing tool for the Property.

In the current version of the specification the rules for editing Properties are as follows:

- A service MAY impose validation rules on the values of specific types of Properties
- If a client attempts to set a Property to an invalid value, then the service MAY reject the change
- Where possible, the validation rules for a type of Property SHOULD be defined in the Property description

Future versions of the VOSpace specification may extend the PropertyDescription to include more specific machine readable validation rules for a Property type.

Note that at the time of writing, the schema for registering validation rules in PropertyDescriptions has not been finalized.

## 3.2.4 Standard properties

Property URIs and PropertyDescriptions for the core set of Properties are registered under a StandardKeyEnumeration resource [VOStd] in the IVOA registry with the resource identifier ivo://ivoa.net/vospace/core. The following URIs SHOULD be used to represent the service properties:

- ivo://ivoa.net/vospace/core#title SHALL be used as the property URI denoting a name given to the resource
- ivo://ivoa.net/vospace/core#creator SHALL be used as the property URI denoting an entity primarily responsible for making the resource
- ivo://ivoa.net/vospace/core#subject SHALL be used as the property URI denoting the topic of the resource
- ivo://ivoa.net/vospace/core#description SHALL be used as the property URI denoting an account of the resource
- ivo://ivoa.net/vospace/core#publisher SHALL be used as the property URI denoting an entity responsible for making the resource available
- ivo://ivoa.net/vospace/core#contributor SHALL be used as the property URI denoting an entity responsible for making contributions to this resource
- ivo://ivoa.net/vospace/core#date SHALL be used as the property URI denoting a point or period of time associated with an event in the lifecycle of the resource
- ivo://ivoa.net/vospace/core#type SHALL be used as the property URI denoting the nature or genre of the resource
- ivo://ivoa.net/vospace/core#format SHALL be used as the property URI denoting the file format, physical medium, or dimensions of the resource
- ivo://ivoa.net/vospace/core#identifier SHALL be used as the property URI denoting an unambiguous reference to the resource within a given context
- ivo://ivoa.net/vospace/core#source SHALL be used as the property URI denoting a related resource from which the described resource is derived
- ivo://ivoa.net/vospace/core#language SHALL be used as the property URI denoting a language of the resource
- ivo://ivoa.net/vospace/core#relation SHALL be used as the property URI denoting a related resource
- ivo://ivoa.net/vospace/core#coverage SHALL be used as the property URI denoting the spatial or temporal topic of the resource, the spatial applicability of the resource, or the jurisdiction under which the resource is relevant
- ivo://ivoa.net/vospace/core#rights SHALL be used as the property URI denoting information about rights held in and over the resource
- ivo://ivoa.net/vospace/core#availableSpace SHALL be used as the property URI denoting the amount of space available within a container
- ivo://ivoa.net/vospace/core#groupread SHALL be used as the property URI denoting the list of groups which can only read this resource (read-only)
- ivo://ivoa.net/vospace/core#groupwrite SHALL be used as the property URI denoting the list of groups which can read and write to this resource (read-write)
- ivo://ivoa.net/vospace/core#publicread SHALL be used as the property URI denoting whether this resource is world readable (anon-read-only)
- ivo://ivoa.net/vospace/core#quota SHALL be used as the property URI denoting the value of a system quota on the resource
- ivo://ivoa.net/vospace/core#length SHALL be used as the property URI denoting the length or size of a resource
- ivo://ivoa.net/vospace/core#mtime SHALL be used as the property URI denoting the data modification time
- ivo://ivoa.net/vospace/core#ctime SHALL be used as the property URI denoting status change (aka metadata modification) time
- ivo://ivoa.net/vospace/core#btime SHALL be used as the property URI denoting initial creation time

However, this is not intended to be a closed list, different implementations are free to define and use their own Properties.

## 3.3 Capabilities

A Capability describes a third-party interface to a node. One application of this would be to enable data access to a node or its contents using a 3rd party service interface.

A Capability representation SHALL have the following members:

- uri: the Capability identifier
- endpoint: the endpoint URL to use for the third-party interface
- param: a set of parameters for the capability

### 3.3.1 Example use cases

A ContainerNode containing image files may offer a DAL SIAP capability so that the images in the container can be accessed using a SIAP service. In this way, a user could create a (DAL enabled) Container in VOSpace, transfer some images into it and then query the set of images using the SIAP interface.

Another example is a DataNode that provides an iRODS capability so that the data replication for this data object can be handled using the iRODS service API.

### 3.3.2 Capability identifiers

Every new type of Capability SHALL require a unique URI to identify the Capability. The rules for the Capability identifiers are similar to the rules for namespace URIs in XML schema. The only restriction is that it SHALL be a valid (unique) URI.

- An XML schema namespace identifier can be just a simple URN, e.g. urn:my-namespace
- Within the IVOA, the convention for namespace identifiers is to use a HTTP URL pointing to the namespace schema, or a resource describing it.

The VOSpace schema defines Capability identifiers as anyURI. The only restriction is that it SHALL be a valid (unique) URI.

- A Capability URI can be a simple URN, e.g. urn:my-capability

This may be sufficient for testing and development on a private system, but it is not suitable for use on a public service. For a production system, any new Capabilities SHOULD have unique URIs that can be resolved into a description of the Capability. Ideally, these SHOULD be IVO registry URIs that point to a description registered in the IVO registry:

- ivo://my-registry/vospace/capabilities#my-capability

Using an IVO registry URI to identify Capabilities has two main advantages:

- IVO registry URIs are by their nature unique, which makes it easy to ensure that different teams do not accidentally use the same URI
- If the IVO registry URI points to a description registered in the IVO registry, this provides a mechanism to discover how to use the Capability.

### 3.3.3 Capability descriptions

If the URI for a particular Capability is resolvable, i.e. an IVO registry identifier or a HTTP URL then it SHOULD point to an XML resource that describes the Capability.

A CapabilityDescription SHOULD describe the third-party interface and how it should be used in this context. A CapabilityDescription SHOULD have the following members:

- *uri*: the formal URI of the Capability
- *DisplayName*: a simple display name of the Capability.
- *Description*: a text block describing the third-party interface and how it should be used in this context.

Note that at the time of writing, the schema for registering CapabilityDescriptions in the IVO registry has not been finalized.

### 3.3.4 UI display name

If a client is unable to resolve a Capability identifier into a description then it may just display the identifier as a text string:

- Access data using urn:edu.sdsc.irods

If a client can resolve the Capability identifier into a description then the client may use the information in the description to display a human readable name and description of the Capability:

- Access data using iRODS

### 3.3.5 Standard capabilities

Capability URIs and CapabilityDescriptions for the core set of Capabilities are registered under a StandardKeyEnumeration resource [VOStd] in the IVOA registry with the resource identifier ivo://ivoa.net/vospace/core.. The following URIs SHALL be used to represent the service capabilities:

- ivo://ivoa.net/vospace/core#vospace-1.0 SHALL be used as the capability URI for a VOSpace 1.0 service
- ivo://ivoa.net/vospace/core#vospace-1.1 SHALL be used as the capability URI for a VOSpace 1.1 service
- ivo://ivoa.net/vospace/core#vospace-2.0 SHALL be used as the capability URI for a VOSpace 2.0 service

If a service implementation supports more than one version of the VOSpace interface then these capability URIs can be used with a VOSpace service to identify different VOSpace capabilities for a node.

One use case for this would be a VOSpace 1.1 client talking to a service that implements both VOSpace 1.0 and VOSpace 1.1, where the client is acting on behalf of a third party agent that only understands VOSpace 1.0. In this case, the client can use the information in the VOSpace 1.0 capability to direct the third party agent to the VOSpace 1.0 endpoint.

Other standard service interfaces will also be registered, e.g.

- Cone Search
- SIAP
- SSAP
- TAP

However, this is not intended to be a closed list and different implementations are free to define and use their own Capabilities.

## 3.4 Views

A View describes the data formats and contents available for importing or exporting data to or from a VOSpace node.

The metadata for a DataNode contains two lists of Views.

- *accepts* is a list of Views that the service can accept for importing data into the Node
- *provides* is a list of Views that the service can provide for exporting data from Node

A View representation SHALL have the following members:

- *uri*: the View identifier
- *original*: an optional boolean flag to indicate that the View preserves the original bit pattern of the data
- *param*: a set of name-value pairs that can be used to specify additional arguments for the View

### 3.4.1 Example use cases

**Simple file store**

A simple VOSpace system that stores data as a binary files can just return the contents of the original file. The client supplies a View identifier when it imports the data, and the service uses this information to describe the data to other clients.

A file based system can use the special case identifier 'ivo://ivoa.net/vospace/core#viewany' to indicate that it will accept any data format or View for a Node.

For example:

- A client imports a file into the service, specifying a View to describe the file contents
- The service stores the data as a binary file and keeps a record of the View
- The service can then use the View supplied by the client to describe the data to other clients

This type of service is not required to understand the imported data, or to verify that it contents match the View, it treats all data as binary files.

*3.4.1.1 Database store*

A VOSpace system that stores data in database tables would need to be able to understand the data format of an imported file in order to parse the data and store it correctly. This means that the service can only accept a specific set of Views, or data formats, for importing data into the Node.

In order to tell the client what input data formats it can accept, the service publishes a list of specific Views in the accepts list for each Node.

On the output side, a database system would not be able to provide access to the original input file. The contents of file would have been transferred into the database table and then discarded. The system has to generate the output results from the contents of the database table.

In order to support this, the service needs to be able to tell the client what Views of the data are available.

A database system may offer access to the table contents as either VOTable or FITS files, it may also offer zip or tar.gz compressed versions of these. In which case the system needs to be able to express nested file formats such as 'zip containing VOTable' and 'tar.gz containing FITS'.

A service may also offer subsets of the data. For example, a work flow system may only want to look at the table headers to decide what steps are required to process the data. If the table contains a large quantity of data, then downloading the whole contents just to look at the header information is inefficient. To make this easier, a database system may offer a 'metadata only' View of the table, returning a VOTable or FITS file containing just the metadata headers and no rows.

So our example service may want to offer the following Views of a database table:

- Table contents as FITS
- Table contents as VOTable
- Table contents as zip containing FITS
- Table contents as zip containing VOTable
- Table contents as tar.gz containing FITS
- Table contents as tar.gz containing VOTable
- Table metadata as FITS
- Table metadata as VOTable

The service would publish this information as a list of Views in the provides section of the metadata for each Node.

The VOSpace specification does not mandate what Views a service must provide. The VOSpace specification is intended to provide a flexible mechanism enabling services to describe a variety of different Views of data. It is up to the service implementation to decide what Views of the data it can accept and provide.

### 3.4.2 View identifiers

Every new type of View SHALL require a unique URI to identify the View and its content.

The rules for the View identifiers are similar to the rules for namespace URIs in XML schema. The only restriction is that it SHALL be a valid (unique) URI.

- An XML schema namespace identifier can be just a simple URN, e.g. urn:mynamespace
- Within the IVOA, the convention for namespace identifiers is to use a HTTP URL pointing to the namespace schema, or a resource describing it

The current VOSpace schema defines View identifiers as anyURI. The only restriction is that it SHALL be a valid (unique) URI.

- A View URI can be a simple URN, e.g. urn:myview

This may be sufficient for testing and development on a private system, but it is not scalable for use on a public service.

For a production system, any new Views SHOULD have unique URIs that can be resolved into to a description of the View.

Ideally, these should be IVO registry URIs that point to a description registered in the IVO registry:

- ivo://myregistry/vospace/views#myview

Using an IVO registry URI to identify Views has two main advantages :

- IVO registry URIs are by their nature unique, which makes it easy to ensure that different teams do not accidentally use the same URI
- If the IVO registry URI points to a description registered in the IVO registry, this provides a mechanism to discover what the View contains

### 3.4.3 View descriptions

If the URI for a particular View is resolvable, i.e. an IVO registry identifier or a HTTP URL, then it SHOULD point to an XML resource that describes the View.

A ViewDescription SHOULD describe the data format and/or content of the view.

A ViewDescription SHOULD have the following members :

- *Uri*: the formal URI of the View
- *DisplayName*: A simple text display name of the View
- *Description*: Text block describing the data format and content of the View

A ViewDescription MAY have the following optional members:

- MimeType : the standard MIME type of the View, if applicable

However, at the time of writing, the schema for registering ViewDescriptions in the IVO registry has not been finalized.

#### 3.4.3.1 UI display name

If a client is unable to resolve a View identifier into a description, then it MAY just display the identifier as a text string:

- Download as urn:table.meta.fits

If the client can resolve the View identifier into a description, then the client MAY use the information in the description to display a human readable name and description of the View:

- Download table metadata as FITS header

#### 3.4.3.2 MIME types

If a VOSpace service provides HTTP access to the data contained in a Node, then if the ViewDescription contains a MimeType field, this SHOULD be included in the appropriate header field of the HTTP response.

### 3.4.4 Default views

The following standard URIs are registered under a StandardKeyEnumeration resource [VOStd] in the IVOA registry with the resource identifier ivo://ivoa.net/vospace/core. They SHALL be used to refer to the default import and export views:

- ivo://ivoa.net/vospace/core#anyview SHALL be used as the view URI to indicate that a service will accept any view for an import operation
- ivo://ivoa.net/vospace/core#binaryview SHALL be used as the view URI to import or export data as a binary file
- ivo://ivoa.net/vospace/core#defaultview SHALL be used by a client to indicate that the service should choose the most appropriate view for a data export

#### 3.4.4.1 Default import view

If a client imports data using this view, the data SHALL be treated as a binary BLOB, and stored as is with no additional processing. This is equivalent to the application/binary MIME type.

Note, this view is OPTIONAL, and the service may throw a ViewNotSupported exception if it does not accept this view. In particular, this view cannot be used to import data into a StructuredDataNode as the service needs to know about and understand the data format to be able to create the StructuredDataNode.

Note, this view is only valid for the data import operations, pullToVoSpace and pushToVoSpace. If this view is requested in an export operation, pullFromVoSpace and pushToVoSpace, then the service SHOULD throw a ViewNotSupported exception.

*3.4.4.2 Default export view*

If a client requests data using this view, the server SHALL choose whichever of the available views (the server) thinks is the most appropriate, based on how the data is stored. In a simple file-based system, this will probably be the same format that the data was originally imported in. In a database table system, this will probably either be VOTable or CVS, depending on the level of metadata available.

Note, this view is OPTIONAL, and the server may throw a ViewNotSupported exception if it does not provide this view. However, in most cases, it is expected that a service would be able to select an appropriate 'default' format for data held within the service.

Note, this view is only valid for the data export operations, pullFromVoSpace and pushFromVoSpace. If this view is requested in an import operation, pullToVoSpace and pushToVoSpace, then the service SHOULD throw a ViewNotSupported fault.

### 3.4.5 Container views

In VOSpace 2.0, a view of a ContainerNode describes the data representation (format) of a file or data stream that represents the combined contents of the node and its children. If the view describes an archive format (tar, zip, etc.) then a service that accepts this view (format) for a ContainerNode SHALL provide access to the archive contents as children nodes of the container. Whether or not the service actually unpacks the archive is implementation dependent but the service SHALL behave as though it has done so. For example, a client may want to upload a tar file containing several images to a VOSpace service. If they associate it with (upload it to) a (Un)structuredDataNode then it will be treated as a blob and its contents will not be available. However, if they upload it to a ContainerNode with an accepts view of "tar" then the image files within the tar file will be represented as children nodes of the ContainerNode and accessible like any other data object within the space.

If a service provides an archive format (tar, zip, etc.) view of a ContainerNode then the service SHALL package the contents of the container and all its child nodes in the specified format.

## 3.5 Protocols

A Protocol describes the parameters required to perform a data transfer using a particular protocol.

A Protocol representation SHALL have the following members:

- *uri*: the Protocol identifier
- *endpoint*: the endpoint URL to use for the data transfer
- *param*: A list of name-value pairs that specify any additional arguments required for the transfer

### 3.5.1 Protocol identifiers

Every new type of Protocol requires a unique URI to identify the Protocol and how to use it.

The rules for the Protocol identifiers are similar to the rules for namespace URIs in XML schema. The only restriction is that it SHALL be a valid (unique) URI

- An XML schema namespace identifier can be just a simple URN, e.g. urn:mynamespace
- Within the IVOA, the convention for namespace identifiers is to use a HTTP URL pointing to the namespace schema, or a resource describing it

The current VOSpace schema defines Protocol identifiers as anyURI. The only restriction is that it SHALL be a valid (unique) URI.

- A Protocol URI can be a simple URN, e.g. urn:myprotocol

This may be sufficient for testing and development on a private system, but it is not scalable for use on a public service.

For a production system, any new Protocols SHOULD have unique URIs that can be resolved into to a description of the Protocol.

Ideally, these should be IVO registry URIs that point to a description registered in the IVO registry :

- ivo://myregistry/vospace/protocols#myprotocol

Using an IVO registry URI to identify Protocols has two main advantages:

- IVO registry URIs are by their nature unique, which makes it easy to ensure that different teams do not accidentally use the same URI
- If the IVO registry URI points to a description registered in the IVO registry, this provides a mechanism to discover how to use the Protocol

### 3.5.2 Protocol descriptions

If the URI for a particular Protocol is resolvable, i.e. an IVO registry identifier or a HTTP URL, then it SHOULD point to an XML resource that describes the Protocol.

A ProtocolDescription SHOULD describe the underlying transport protocol, and how it should be used in this context.

A ProtocolDescription SHOULD have the following members:

- *uri*: the formal URI of the Protocol
- *DisplayName*: A simple display name of the Protocol
- *Description*: Text block describing describing the underlying transport protocol, and how it should be used in this context

However, at the time of writing, the schema for registering ProtocolDescriptions in the IVO registry has not been finalized.

### 3.5.2.1 UI display name

If a client is unable to resolve a Protocol identifier into a description, then it MAY just display the identifier as a text string:

- Download using urn:myprotocol

If the client can resolve the Protocol identifier into a description, then the client MAY use the information in the description to display a human readable name and description of the Protocol:

- Download using standard HTTP GET

### 3.5.3 Standard protocols

Protocol URIs and ProtocolDescriptions for the core set of standard transport protocols are registered under a StandardKeyEnumeration resource [VOStd] in the IVOA registry with the resource identifier ivo://ivoa.net/vospace/core. The following URIs SHALL be used to represent the standard protocols:

- ivo://ivoa.net/vospace/core#httpget SHALL be used as the protocol URI for a HTTP GET transfer
- ivo://ivoa.net/vospace/core#httpput SHALL be used as the protocol URI for a HTTP PUT transfer

However, this is not intended to be a closed list, different implementations are free to define and use their own transfer Protocols.

### 3.5.4 Custom protocols

There are several use cases where a specific VOSpace implementation may want to define and use a custom VOSpace transfer Protocol, either extending an existing Protocol, or defining a new one.

### 3.5.4.1 SRB gateway

One example would be a VOSpace service that was integrated with a SRB system. In order to enable the service to use the native SRB transport protocol to transfer data, the service providers would need to register a new ProtocolDescription to represent the SRB transport protocol.

The ProtocolDescription would refer to the technical specification for the SRB transport protocol, and define how it should be used to transfer data to and from the VOSpace service.

Clients that do not understand the SRB transport protocol would not recognize the URI for the SRB Protocol, and would ignore Transfer options offered using this Protocol.

Clients that were able to understand the SRB transport protocol would recognize the URI for the SRB Protocol, and could use the 'srb://..' endpoint address in a Protocol option to transfer data using the SRB transport protocol.

Enabling different groups to define, register and use their own custom Protocols in this way means that support for new transport protocols can be added to VOSpace systems without requiring changes to the core VOSpace specification.

In this particular example, it is expected that one group within the IVOA will work with the SRB team at SDSC to define and register the Protocol URI and ProtocolDescription for using the SRB protocol to transfer data to and from VOSpace systems.

Other implementations that plan to use the SRB transport protocol in the same way could use the same Protocol URI and ProtocolDescription to describe data transfers using the SRB transport protocol.

The two implementations would then be able use the common Protocol URI to negotiate data transfers using the SRB transport protocol.

### 3.5.4.2 Local NFS transfers

Another example of a custom Protocol use case would to transfer data using the local NFS file system within an institute.

If an institute has one or more VOSpace services co-located with a number of data processing applications, all located within the same local network, then it would be inefficient to use HTTP GET and PUT to transfer the data between the services if they could all access the same local file system.

In this case, the local system administrators could register a custom ProtocolDescription which described how to transfer data using their local NFS file system.

- ivo://my.institute/vospace/protocols#internalnfs

Data transfers using this Protocol would be done using file:// URLs pointing to locations within the local NFS file system:

- file:///net/host/path/file

These URLs would only have meaning to services and applications located within the local network, and would not be usable from outside the network.

Services and applications located within the local network would be configured to recognize the custom Protocol URI, and to use local file system operations to move files within the NFS file system.

Services and applications located outside local network would not recognize the custom Protocol URI and so would not attempt to use the internal file URLs to transfer data.

Note that in this example the custom Protocol URI and the associated ProtocolDescription refer to data transfers using file URLs within a specific local NFS file system.

If a different institute wanted to use a similar system to transfer data within their own local network, then they would have to register their own custom Protocol URI and associated ProtocolDescription.

The two different Protocol URIs and ProtocolDescriptions describe how to use the same underlying transport protocol (NFS) in different contexts.

Enabling different groups to define, register and use their own custom Protocols in this way means that systems can be configured to use the best available transport protocols for transferring data, without conflicting with other systems who may be using similar a transport protocol in a different context.

## 3.6 Transfers

A Transfer describes the details of a data transfer to or from a space.

A UWS Job representation [UWS] of a Transfer SHALL have the following parameters:

- *target*: denotes the VOSpace node to/from which data is to be transferred
- *direction*: denotes the direction of a data transfer
  - It can be a URI for internal data transfers (move and copy operations) or one of: pushToVoSpace, pullToVoSpace, pushFromVoSpace or pullFromVoSpace - for an external data transfer.
- *view*: denotes the requested View
  - For the transfer to be valid, the specified View must match one of those listed in the accepts or provides list of the Node.
  - For an internal data transfer, this parameter is not required.
- *protocol*: denotes the transfer protocol(s) to use
  - A transfer may contain more than one protocol with different Protocol URIs.
  - For an internal data transfer, this parameter is not required.
- *keepBytes*: denotes whether the source object is to be kept in an internal data transfer, i.e. distinguishes between a move and a copy
  - For an external data transfer, this parameter is not required.

This representation will be used as a child of the *jobInfo* element in a UWS Job representation.

The representation of the results of a Transfer SHALL have the following members:

- *target*: denotes the VOSpace node to/from which data is to be transferred
- *direction*: denotes the direction of a data transfer
  - It can be a URI for internal data transfers (move and copy operations) or one of: pushToVoSpace, pullToVoSpace, pushFromVoSpace or pullFromVoSpace - for an external data transfer.
- *view*: A View specifying the requested View
  - For the transfer to be valid, the specified View must match one of those listed in the accepts or provides list of the Node
  - For an internal data transfer, this parameter is not required.
- *protocol*: one or more Protocols specifying the transfer protocols to use
  - A Transfer may contain more than one Protocol element with different Protocol URIs
  - A Transfer may contain more than one Protocol element with the same Protocol URI with different endpoints
  - For an internal data transfer, this parameter is not required.

### 3.6.1 Service-initiated transfers

Two of the external data transfers (pullToVoSpace and pushFromVoSpace) rely on the server performing the data transfer itself.

The client constructs a Job request containing details of the View and one or more Protocol elements with valid endpoint addresses.

The service MAY ignore Protocols with URIs that it does not recognize.

If the server is unable to handle any of the requested Protocols in a Transfer request, then it SHALL respond with a fault.

The order of the Protocols in the request indicates the order of preference that the client would like the server to use. However, this is only a suggestion, and the server is free to use its own preferences to select which Protocol it uses first.

The service selects the first Protocol it wants to use from the list and attempts to transfer the data using the Protocol endpoint.

If the first attempt fails, the server may choose another Protocol from the list and re-try the transfer using that Protocol endpoint. The status flag will be updated to reflect this.

The server may attempt to transfer the data using any or all of the Protocols in the list until either, the data transfer succeeds, or there are no more Protocol options left.

The server SHALL be allowed to only use each Protocol option once. This allows a data source to issue one time URLs for a Transfer, and cancel each URL once it has been used.

Once one of the Protocol options succeeds the transfer SHALL be considered to have completed - the status flag needs to be updated to reflect this -, and the server is not allowed to use any of the remaining Protocol options. This allows a data source to issue a set of one time URLs for a transfer, and to cancel any unused URLs once the transfer has been completed.

Some Protocols MAY require the service to call a callback address when a data transfer completes. This behavior is specific to the Protocol, and SHOULD be defined in the ProtocolDescription.

If none of the Protocol options succeed, then the transfer SHALL be considered to have failed, and the service SHALL return a fault containing details of the Protocol options it tried. The status flag will be updated to reflect this.

### 3.6.2 Client-initiated transfers

Two of the VOSpace external transfer methods rely on an external actor performing the data transfer outside the scope of the service call.

In these methods, the client sends a Job request to the server which SHOULD contain details of the View and one or more protocol parameters.

In effect, the client is sending a list of Protocols that it (the client) wants to use for the transfer.

The service MAY ignore Protocols with URIs that it does not recognize.

The service selects the Protocols from the request that it is capable of handling, and builds a Transfer results response containing the selected Protocol elements filling in valid endpoint addresses for each of them.

The order of the Protocol elements in the request indicates the order of preference that the client would like to use. However, this is only a suggestion, and the server is free to use its own preferences when generating the list of Protocols in the response.

In effect, the server is responding with the subset of the requested Protocols that it (the server) is prepared to offer.

If the server is unable to accept any of the requested Protocols, then it SHALL respond with a fault.

On receipt of the response, the client can use the list of Protocols itself, or pass them on to another agent to perform the data transfer on its behalf.

The agent MAY ignore Protocols URIs that it does not recognize.

The agent selects the first Protocol it wants to use from the list and attempts to transfer the data using the Protocol endpoint. The status flag will be updated to reflect this.

If the first attempt fails, the agent MAY choose another Protocol from the list and re-try the transfer using that Protocol endpoint.

The agent MAY attempt to transfer the data using any or all of the Protocols in the list until either, the data transfer succeeds, or there are no more Protocol options left.

The agent SHALL be allowed to only use each Protocol option once. This allows a data source to issue one time URLs for a Transfer, and cancel each URL once it has been used.

Once one of the Protocol options succeeds the transfer SHALL be considered to have completed and the status flag will be updated correspondingly. The agent is not allowed to use any of the remaining unused Protocol options. This allows a data source to issue a set of one time URLs for a transfer, and to cancel any unused URLs once the transfer has been completed.

Some Protocols MAY require the agent to call a callback address when a data transfer completes. This behavior is specific to the Protocol, and SHOULD be defined in the ProtocolDescription.

If none of the Protocol options succeed, then the transfer SHALL be considered to have failed and the status will be updated.

## 3.7 Searches

A Search describes the details of data objects in the space which meet specified search criteria, i.e. that are the results of a submitted search request.

A UWS Job representation of a Search SHALL have the following parameters:

- *uri*: An OPTIONAL identifier indicating from which item to continue a search
- *limit*: An OPTIONAL limit indicating the maximum number of results in the response
  - No limit indicates a request for an unpaged list. However the server MAY still impose its own limit on the size of an individual response, splitting the results into more than one page if required
- *detail*: The level of detail in the results
  - min : The list contains the minimum detail for each Node with all optional parts removed - the node type should be returned
    - e.g. <node uri="vos://service/name" xsi:type="Node"/>
  - max : The list contains the maximum detail for each Node, including any xsi:type specific extensions
  - properties : The list contains a basic node element with a list of properties for each Node with no xsi:type specific extensions.
- *matches*: An OPTIONAL search string consisting of properties and values to match against and joined in conjunction (and) or disjunction (or).
  - Each property-value pair consists of the uri identifying a particular property and a regular expression against which the property values are to be matched: 'uri' = 'regex'
  - The match pairs can be combined in conjunction and/or disjunction using 'and' and 'or' respectively. For example: "(property1 = 'value1' and property2 = 'value2') or property3 = 'value3'".
  - The regex syntax SHALL comply with POSIX conventions.
- *node* : An OPTIONAL URI(s) identifying the starting node for a search, i.e., the *matches* constraints are applied to this node and its children

This representation will be used as a child of the *jobInfo* element in a UWS Job representation. For example:

```
<uws:jobInfo>
  <vos:search>
    <vos:detail>properties</vos:detail>
    <vos:matches>ivo://ivoa.net/vospace/core#description='galax'</vos:matches>
```

```
        <vos:search>
    <uws:jobInfo>
```

The representation of the results of a Search SHALL have the following members:

- *nodes*: A list containing zero or more Nodes of appropriate detail identifying the target URIs meeting the search criteria

## 3.8 REST bindings

In a REST (Representational State Transfer) binding of VOSpace, each of the objects defined below is available as a web resource with its own URI.

| | |
|---|---|
| /properties | the properties employed in the space |
| /views | the views employed in the space |
| /protocols | the protocols employed in the space |
| /searches | the searches of the space |
| /nodes/(node-path) | a Node under the nodes of the space |
| /transfers | the transfers for the space |
| /transfers/(job-id)/results/transferDetails | the transfer details for synchronous jobs |

The service implementor is free to choose the names given in parentheses above; the other names are part of the VOSpace 2.0 standard.

The endpoint /sync is also defined to receive HTTP POST requests of synchronous transfer jobs. This should respond with a redirect to the transfer details for synchronous jobs resource: http://rest-endpoint/transfers/(jobid)/results/transferDetails. Synchronous transfers are limited to pushToVoSpace and pullFromVoSpace operations only where the client is requesting endpoint URLs where it can read or write data.

In addition, the following Nodes, appearing as part of a node-path, are reserved:

- .auto indicates that the service should auto-generate an endpoint URI to replace this placeholder (Note: that this is an OPTIONAL feature of a service)
- .null indicates an endpoint that discards all data written to it and provides no data when read from, i.e. a bit bucket

The standardID for the service is: ivo://ivoa.net/std/VOSpace/v2.0. Available resources will then just be ivo://ivoa.net/std/VOSpace/v2.0#<resourceName>, e.g., ivo://ivoa.net/std/VOSpace/v2.0#nodes.

# 4 Access Control

[NOTE: use HTTPS with client authentication and valid X.509 certificate]

The access control policy for a VOSpace is defined by the implementor of that space according to the use cases for which the implementation is intended.

A human-readable description of the implemented access policy must be declared in the registry metadata for the space.

These are the most probable access policies:

- No access control is asserted. Any client can create, read, write and delete nodes anonymously
- No authorization is required, but clients must authenticate an identity (for logging purposes) in each request to the space
- Clients may not create or change nodes (i.e. the contents of the space are fixed by the implementation or set by some interface other than VOSpace), but any client can read any node without authentication
- Nodes are considered to be owned by the user who created them. Only the owner can operate on a node

No operations to modify the access policy (e.g. to set permissions on an individual node) are included in this standard. Later versions may add such operations.

Where the access policy requires authentication of callers, the VOSpace service SHALL support one of the authentication methods defined in the IVOA Single Sign On profile.

# 5 Web service operations

A VOSpace 2.0 service SHALL be a REST service with the following operations:

Note: The URL http://(rest-endpoint)/nodes denotes the URL of the top node in the VOSpace. The URL http://(rest-endpoint)/nodes/(path) denotes a specific node within the VOSpace.

Note: When representing a Fault, the exact specified fault name SHALL be used. If this is followed by some details, the fault SHALL be separated from the subsequent characters by whitespace.

Note: If the service is able to detect an internal system failure (associated with an HTTP 500 status code) then it should indicate this as described below if possible.

## 5.1 Service metadata

These operations return comprehensive lists of service-level metadata, e.g. all protocols supported by the service. Individual nodes within a service, however, may not necessarily support all of these, i.e. only container nodes may support archive formats such as zip or tar.

### 5.1.1 getProtocols

Get a list of the transfer Protocols supported by the space service

*5.1.1.1 Request*

- A HTTP GET to http://rest-endpoint/protocols

*5.1.1.2 Response*

- A HTTP 200 status code
- A Protocols representation giving:
  - accepts: A list of Protocols that the service SHALL accept
    - In this context 'accepting a protocol' means that the service SHALL act as a client for the protocol
    - e.g. 'accepting httpget' means the service can read data from an external HTTP web server
  - provides: A list of Protocols that the service SHALL provide
    - In this context 'providing a protocol' means that the service SHALL act as a server for the protocol
    - e.g. 'providing httpget' means the service can act as a http web server

*5.1.1.3 Faults*

- The service SHOULD throw a HTTP 500 status code including an InternalFault fault in the entity body if the operation fails

```
Example: getProtocols

  > curl -v http://localhost:8000/vospace-2.0/protocols

* Connected to localhost (127.0.0.1) port 8000 (#0)
> GET /protocols HTTP/1.1
> User-Agent: curl/7.19.4 (universal-apple-darwin10.0) libcurl/7.19.4 OpenSSL/0.9.8l zlib/1.2.3
> Host: localhost:8000
> Accept: */*
>
< HTTP/1.1 200 OK
< Date: Tue, 09 Mar 2010 04:59:12 GMT
< Content-Length: 309
< Content-Type: text/html
< Allow: GET, HEAD, POST, PUT
< Server: CherryPy/3.1.2
<
<protocols xmlns="http://www.ivoa.net/xml/VOSpace/v2.0">
   <accepts>
      <protocol uri="ivo://ivoa.net/vospace/core#httpget"/>
      <protocol uri="ivo://ivoa.net/vospace/core#httpput"/>
   </accepts>
   <provides>
      <protocol uri="ivo://ivoa.net/vospace/core#httpget"/>
      <protocol uri="ivo://ivoa.net/vospace/core#httpput"/>
   </provides>
</protocols>
* Connection #0 to host localhost left intact
* Closing connection #0
```

### 5.1.2 getViews

Get a list of the *Views* and data formats supported by the space service

*5.1.2.1 Request*

- A HTTP GET to http://rest-endpoint/views

*5.1.2.2 Response*

- A HTTP 200 status code
- A Views representation giving:
  - *accepts*: A list of Views that the service SHALL accept
    - In this context 'accepting a view' means that the service SHALL receive input data in this format
    - A simple file based system may use the reserved View URI ivo://net.ivoa.vospace/views/any to indicate that it can accept data in any format
  - *provides*: A list of Views that the service SHALL provide
    - In this context 'providing a view' means that the service SHALL produce output data in this format
    - A simple file based system may use the reserved View URI ivo://net.ivoa.vospace/views/any to indicate that it can provide data in any format

*5.1.2.3 Faults*

- The service SHOULD throw a HTTP 500 status code including an InternalFault fault in the entity body if the operation fails

---

**Example: getViews**

```
> curl -v "http://localhost:8000/views"

* Connected to localhost (127.0.0.1) port 8000 (#0)
> GET /views HTTP/1.1
> User-Agent: curl/7.19.4 (universal-apple-darwin10.0) libcurl/7.19.4 OpenSSL/0.9.8l zlib/1.2.3
> Host: localhost:8000
> Accept: */*
>
< HTTP/1.1 200 OK
< Date: Tue, 09 Mar 2010 19:36:17 GMT
< Content-Length: 181
< Content-Type: text/html
< Allow: GET, HEAD, POST, PUT
< Server: CherryPy/3.1.2
<
<views xmlns="http://www.ivoa.net/xml/VOSpace/v2.0">
  <accepts>
    <view uri="ivo://ivoa.net/vospace/core#anyview"/>
  </accepts>
  <provides>
    <view uri="ivo://ivoa.net/vospace/core#defaultview"/>
  </provides>
</views>
* Connection #0 to host localhost left intact
* Closing connection #0
```

---

### 5.1.3 getProperties

#### 5.1.3.1 Request

- A HTTP GET to http://rest-endpoint/properties

#### 5.1.3.2 Response

- A HTTP 200 status code
- A Properties representation including:
  - *accepts*: A list of identifiers for the Properties that the service SHALL accept and understand. This refers to metadata (Properties) that is implementation dependent but can be used by a client to control operational aspects of the service: for example, a VOSpace implementation might allow individual users to control the permissions on data objects via a Property called "permissions". If the VOSpace receives a data object with this Property then it 'understands' what this property refers to and can deal with it accordingly.
  - *provides*: A list of identifiers for the Properties that the service SHALL provide
  - *contains*: A list of identifiers for all the Properties currently used by Nodes within the service

#### 5.1.3.3 Faults

- The service SHOULD throw a HTTP 500 status code including an InternalFault fault in the entity body if the operation fails

---

**Example: getProperties**

```
> curl -v "http://localhost:8000/properties"

* Connected to localhost (127.0.0.1) port 8000 (#0)
> GET /properties HTTP/1.1
> User-Agent: curl/7.19.4 (universal-apple-darwin10.0) libcurl/7.19.4 OpenSSL/0.9.8l zlib/1.2.3
> Host: localhost:8000
> Accept: */*
>
< HTTP/1.1 200 OK
< Date: Tue, 09 Mar 2010 19:43:23 GMT
< Content-Length: 644
< Content-Type: text/html
< Allow: GET, HEAD, POST, PUT
< Server: CherryPy/3.1.2
<
<properties xmlns="http://www.ivoa.net/xml/VOSpace/v2.0">
  <accepts>
    <property uri="ivo://ivoa.net/vospace/core#title"/>
    <property uri="ivo://ivoa.net/vospace/core#creator"/>
    <property uri="ivo://ivoa.net/vospace/core#description"/>
  </accepts>
  <provides>
    <property uri="ivo://ivoa.net/vospace/core#availableSpace"/>
    <property uri="ivo://ivoa.net/vospace/core#httpput"/>
  </provides>
  <contains>
    <property uri="ivo://ivoa.net/vospace/core#availableSpace"/>
    <property uri="ivo://ivoa.net/vospace/core#title"/>
    <property uri="ivo://ivoa.net/vospace/core#creator"/>
    <property uri="ivo://ivoa.net/vospace/core#description"/>
  </contains>
</properties>
```

```
* Connection #0 to host localhost left intact
* Closing connection #0
```

## 5.2 Creating and manipulating data nodes

### 5.2.1 createNode

Create a new node at a specified location

*5.2.1.1 Request*

- A HTTP PUT of a node representation to the node URL:
  - *node*: A template Node (as defined in Section 3.1) for the node to be created

A valid uri attribute SHALL be required. The name .auto is a reserved URI to indicate an auto-generated URI for the destination, i.e. given the following URI vos://service/path/.auto a service SHALL create a new unique URI for the node within vos://service/path.

If the Node xsi:type is not specified then a generic node of type Node is implied.

The permitted values of xsi:type are:

- vos:Node
- vos:DataNode
- vos:UnstructuredDataNode
- vos:StructuredDataNode
- vos:ContainerNode
- vos:LinkNode

When creating a new Node the service MAY substitute a valid subtype, i.e. If xsi:type is set to vos:DataNode then this may be implemented as a DataNode, StructuredDataNode or an UnstructuredDataNode.

The properties of the new Node can be set by adding Properties to the template. Attempting to set a Property that the service considers to be 'readOnly' SHALL cause a PermissionDenied fault. The accepts and provides lists of Views for the Node cannot be set using this method.

The capabilities list for the Node cannot be set using this method.

*5.2.1.2 Response*

- A HTTP 201 status code
- A node representation including
  - *node*: details of the new Node

The *accepts* list of Views for the Node SHALL be filled in by the service based on service capabilities.

The *provides* list of Views for the Node MAY not be filled in until some data has been imported into the Node.

The *capabilities* list for the Node MAY not be filled in until some data has been imported into the Node.

*5.2.1.3 Faults*

- The service SHOULD throw a HTTP 500 status code including an InternalFault fault in the entity body if the operation fails
- The service SHALL throw a HTTP 409 status code including a DuplicateNode fault in the entity body if a Node already exists with the same URI
- The service SHALL throw a HTTP 400 status code including an InvalidURI fault in the entity body if the requested URI is invalid
- The service SHALL throw a HTTP 400 status code including a TypeNotSupported fault in the entity body if the type specified in xsi:type is not supported
- The service SHALL throw a HTTP 401 status code including PermissionDenied fault in the entity body if the user does not have permissions to perform the operation
- If a parent node in the URI path does not exist then the service SHALL throw a HTTP 404 status code including a ContainerNotFound fault in the entity body.
  - For example, given the URI path /a/b/c, the service must throw a HTTP 404 status code including a ContainerNotFound fault in the entity body if either /a or /a/b do not exist.
- If a parent node in the URI path is a LinkNode, the service SHALL throw a HTTP 400 status code including a LinkFound fault in the entity body.
  - For example, given the URI path /a/b/c, the service must throw a HTTP 400 status code including a LinkFound fault in the entity body if either /a or /a/b are LinkNodes.

---

**Example: createNode**

The node to be created (newNode.xml) is:

```
<node xmlns:xsi="http://www.w3.org/2001/XMLSchema-instance" xsi:type="vos:UnstructuredDataNode"
    xmlns="http://www.ivoa.net/xml/VOSpace/v2.0" uri="vos://nvo.caltech!vospace/mydata1">
  <properties>
    <property uri="ivo://ivoa.net/vospace/core#description">My important results</property>
  </properties>
  <accepts/>
  <provides/>
```

```
      <capabilities/>
    </node>

    > curl -v -X PUT -d @newNode.xml "http://localhost:8000/nodes/mydata1"

* Connected to localhost (127.0.0.1) port 8000 (#0)
> PUT /nodes/mydata1 HTTP/1.1
> User-Agent: curl/7.19.4 (universal-apple-darwin10.0) libcurl/7.19.4 OpenSSL/0.9.8l zlib/1.2.3
> Host: localhost:8000
> Accept: */*
> Content-Length: 267
> Content-Type: application/x-www-form-urlencoded
>
< HTTP/1.1 200 Created
< Date: Wed, 10 Mar 2010 00:10:27 GMT
< Content-Length: 323
< Content-Type: text/html
< Allow: GET, HEAD, POST, PUT
< Server: CherryPy/3.1.2
<
* Connection #0 to host localhost left intact
* Closing connection #0
<node xmlns:xsi="http://www.w3.org/2001/XMLSchema-instance"
  xmlns="http://www.ivoa.net/xml/VOSpace/v2.0" xsi:type="vos:UnstructuredDataNode"
  uri="vos://nvo.caltech!vospace/mydata1">
  <properties>
    <property uri="ivo://ivoa.net/vospace/core#description">My important results</property>
  </properties>
  <accepts>
    <view uri="ivo://ivoa.net/vospace/core#anyview"/>
  </accepts>
  <provides/>
  <capabilities/>
</node>
```

### 5.2.2 moveNode

Move a node within a VOSpace service.

Note that this does not cover moving data between two separate VOSpace services.

Moving nodes between separate VOSpace services SHOULD be handled by the client, using the import, export and delete methods.

When the source is a ContainerNode, all its children (the contents of the container) SHALL also be moved to the new destination.

When the destination is an existing ContainerNode, the source SHALL be placed under it (i.e. within the container).

The *Node* type cannot be changed using this method.

#### 5.2.2.1 Request

- A HTTP POST of a Job representation for the transfer (see section 3.6) to http://rest-endpoint/transfers.

.auto is a reserved URI to indicate an autogenerated URI for the destination, i.e. vos://service/path/.auto SHALL cause a new unique URI for the node within vos://service/path to be generated.

#### 5.2.2.2 Response

- The initial response is a HTTP 303 status code with the Location header keyword assigned to the created job.

If an autogenerated URI for the destination was specified in the request then its value SHALL be specified as a result in the Results List for the completed transfer with the id of "destination":

```
<uws:job>
  ...
  <uws:jobInfo>
    <vos:direction>vos://nvo.caltech!vospace/mjg/.auto</vos:direction>
  ...
  <uws:results>
    <uws:result id="destination">vos://nvo.caltech!vospace/mjg/abc123</uws:result>
  </uws:results>
  ...
</uws:job>
```

#### 5.2.2.3 Faults

For all faults, the service shall set the PHASE to "ERROR" in the Job representation. The <errorSummary> element in the Job representation shall be set to the appropriate value for the fault type and the appropriate fault representation (see section 5.5) provided at the error URI: http://rest-endpoint/transfers/(jobid)/error.

| Fault description | errorSummary | Fault representation |
|---|---|---|
| Operation fails | Internal Fault | InternalFault |
| User does not have permissions to perform the operation | Permission Denied | PermissionDenied |

| Source node does not exist | Node Not Found | NodeNotFound |
| Destination node already exists and it is not a ContainerNode | Duplicate Node | DuplicateNode |
| A specified URI is invalid | Invalid URI | InvalidURI |

---

**Example: moveNode**

The Job to be submitted (newJob.xml) is:

```
<vos:transfer xmlns:vos="http://www.ivoa.net/xml/VOSpace/v2.0" >
    <vos:target>vos://nvo.caltech!vospace/mydata1</vos:target>
    <vos:direction>vos://nvo.caltech!vospace/mydata2</vos:direction>
    <vos:keepBytes>false</vos:keepBytes>
</vos:transfer>
```

```
> curl -v -X POST -d @newJob.xml "http://localhost:8000/transfers"
```

```
* Connected to localhost (127.0.0.1) port 8000 (#0)
> POST /transfers HTTP/1.1
> User-Agent: curl/7.19.4 (universal-apple-darwin10.0) libcurl/7.19.4 OpenSSL/0.9.8l zlib/1.2.3
> Host: localhost:8000
> Accept: */*
> Content-Length: 762
> Content-Type: application/x-www-form-urlencoded
>
< HTTP/1.1 303 See Other
< Content-Length: 174
< Server: CherryPy/3.1.2
< Location: http://localhost:8080/transfers/ec200b5ff77641fb841978a85d1f7245
< Allow: GET, HEAD, POST, PUT
< Date: Thu, 11 Mar 2010 19:54:00 GMT
< Content-Type: text/html
<
* Connection #0 to host localhost left intact
* Closing connection #0
This resource can be found at http://localhost:8000/transfers/ec200b5ff77641fb841978a85d1f7245.
```

The status of the job can now be polled at the job location:

```
> curl -v "http://localhost:8000/transfers/ec200b5ff77641fb841978a85d1f7245"
```

```
* Connected to localhost (127.0.0.1) port 8000 (#0)
> GET /transfers/ccfd4ba0dd9f4406b2039c4358ba8ef3 HTTP/1.1
> User-Agent: curl/7.19.4 (universal-apple-darwin10.0) libcurl/7.19.4 OpenSSL/0.9.8l zlib/1.2.3
> Host: localhost:8000
> Accept: */*
>
< HTTP/1.1 200 OK
< Date: Thu, 11 Mar 2010 19:54:02 GMT
< Content-Length: 802
< Content-Type: text/html
< Allow: GET, HEAD, POST, PUT
< Server: CherryPy/3.1.2
<
<uws:job xmlns:uws="http://www.ivoa.net/xml/UWS/v1.0" xmlns:xlink="http://www.w3.org/1999/xlink"
    xmlns:xsi="http://www.w3.org/2001/XMLSchema-instance"
    xmlns:vos="http://www.ivoa.net/xml/VOSpace/v2.0"
    xsi:schemaLocation="http://www.ivoa.net/xml/UWS/v1.0 UWS.xsd ">
  <uws:jobId>ec200b5ff77641fb841978a85d1f7245</uws:jobId>
  <uws:ownerId xsi:nil="true"/>
  <uws:phase>EXECUTING</uws:phase>
  <uws:startTime>2010-03-11T19:54:00.433058</uws:startTime>
  <uws:endTime xsi:nil="true"/>
  <uws:executionDuration>0</uws:executionDuration>
  <uws:destruction xsi:nil="true"/>
  <uws:jobInfo>
    <vos:transfer>
      <vos:target>vos://nvo.caltech!vospace/mydata1</vos:target>
      <vos:direction>vos://nvo.caltech!vospace/mydata2</vos:direction>
      <vos:keepBytes>false</vos:keepBytes>
    </vos:transfer>
  </uws:jobInfo>
  <uws:results/>
* Connection #0 to host localhost left intact
* Closing connection #0
</uws:job>
```

## 5.2.3 copyNode

Copy a node with a VOSpace service.

Note that this does not cover copying data between two separate VOSpace services.

Copying nodes between separate VOSpace services SHOULD be handled by the client, using the import and export methods.

When the source is a ContainerNode, all its children (the full contents of the container) SHALL get copied, i.e. this is a deep recursive copy.

When the destination is an existing ContainerNode, the copy SHALL be placed under it (i.e. within the container).

The *Node* type cannot be changed using this method.

- A HTTP POST of a Job representation for the transfer (see section 3.6) to http://rest-endpoint/transfers.

.auto is the reserved URI to indicate an auto-generated URI for the destination, i.e. vos://service/path/.auto SHALL cause a new unique URI for the node within vos://service/path to be generated.

*5.2.3.2 Response*

- The initial response is a HTTP 303 status code with the Location header keyword assigned to the created job.

If an autogenerated URI for the destination was specified in the request then its value SHALL be specified as a result in the Results List for the completed transfer with the id of "destination":

```
<uws:job>
  ...
  <uws:jobInfo>
     <vos:direction>vos://nvo.caltech!vospace/mjg/.auto</vos:direction>
  ...
  <uws:results>
     <uws:result id="destination">vos://nvo.caltech!vospace/mjg/abc123</uws:result>
  </uws:results>
  ...
</uws:job>
```

*5.2.3.3 Faults*

For all faults, the service shall set the PHASE to "ERROR" in the Job representation. The <errorSummary> element in the Job representation shall be set to the appropriate value for the fault type and the appropriate fault representation (see section 5.5) provided at the error URI: http://rest-endpoint/transfers/(jobid)/error.

| Fault description | errorSummary | Fault representation |
| --- | --- | --- |
| Operation fails | Internal Fault | InternalFault |
| User does not have permissions to perform the operation | Permission Denied | PermissionDenied |
| Source node does not exist | Node Not Found | NodeNotFound |
| Destination node already exists and it is not a ContainerNode | Duplicate Node | DuplicateNode |
| A specified URI is invalid | Invalid URI | InvalidURI |

---

**Example: copyNode**

The Job to be submitted (newJob.xml) is:

```
<vos:transfer xmlns:vos="http://www.ivoa.net/xml/VOSpace/v2.0">
  <vos:target>vos://nvo.caltech!vospace/mydata1</vos:target>
  <vos:direction>vos://nvo.caltech!vospace/mydir/.auto</vos:direction>
  <vos:keepBytes>true</vos:keepBytes>
</vos:transfer>
```

```
> curl -v -X POST -d @newJob.xml "http://localhost:8000/transfers"

* Connected to localhost (127.0.0.1) port 8000 (#0)
> POST /transfers HTTP/1.1
> User-Agent: curl/7.19.4 (universal-apple-darwin10.0) libcurl/7.19.4 OpenSSL/0.9.8l zlib/1.2.3
> Host: localhost:8000
> Accept: */*
> Content-Length: 765
> Content-Type: application/x-www-form-urlencoded
>
< HTTP/1.1 303 See Other
< Content-Length: 174
< Server: CherryPy/3.1.2
< Location: http://localhost:8080/transfers/6fbb28afec94417ba256d705f5e1f966
< Allow: GET, HEAD, POST, PUT
< Date: Thu, 11 Mar 2010 21:28:19 GMT
< Content-Type: text/html
<
* Connection #0 to host localhost left intact
* Closing connection #0
This resource can be found at http://localhost:8000/transfers/6fbb28afec94417ba256d705f5e1f966.
```

The status of the job can now be polled at the job location:

```
> curl -v "http://localhost:8000/transfers/63fa39fb18f244c18c991ed2f96d26cd"

* Connected to localhost (127.0.0.1) port 8000 (#0)
> GET /transfers/63fa39fb18f244c18c991ed2f96d26cd HTTP/1.1
> User-Agent: curl/7.19.4 (universal-apple-darwin10.0) libcurl/7.19.4 OpenSSL/0.9.8l zlib/1.2.3
> Host: localhost:8000
> Accept: */*
>
< HTTP/1.1 200 OK
< Date: Thu, 11 Mar 2010 21:28:21 GMT
< Content-Length: 950
< Content-Type: text/html
< Allow: GET, HEAD, POST, PUT
```

```
< Server: CherryPy/3.1.2
<
<uws:job xmlns:uws="http://www.ivoa.net/xml/UWS/v1.0" xmlns:xlink="http://www.w3.org/1999/xlink"
    xmlns:vos="http://www.ivoa.net/xml/VOSpace/v2.0" xmlns:xsi="http://www.w3.org/2001/XMLSchema-instance"
    xsi:schemaLocation="http://www.ivoa.net/xml/UWS/v1.0 UWS.xsd ">
  <uws:jobId>6fbb28afec94417ba256d705f5e1f966</uws:jobId>
  <uws:ownerId xsi:nil="true"/>
  <uws:phase>COMPLETED</uws:phase>
  <uws:startTime>2010-03-11T21:28:19.200324</uws:startTime>
  <uws:endTime>2010-03-11T21:28:19.200529</uws:endTime>
  <uws:executionDuration>0</uws:executionDuration>
  <uws:destruction xsi:nil="true"/>
  <uws:jobInfo>
    <vos:transfer>
      <vos:keepBytes>true</vos:keepBytes>
      <vos:direction>vos://nvo.caltech!vospace/mydir/.auto</vos:direction>
      <vos:target>vos://nvo.caltech!vospace/mydata1</vos:target>
    </vos:transfer>
  </uws:jobInfo>
  <uws:results>
    <uws:result id="destination" xlink:href="vos://nvo.caltech!vospace/mydir/ef9ce281f5bd4bff92c8991580cddff4"/>
  </uws:results>
* Connection #0 to host localhost left intact
* Closing connection #0
</uws:job>
```

### 5.2.4 deleteNode

Delete a node.

When the target is a ContainerNode, all its children (the contents of the container) SHALL also be deleted.

Note that the same operation can also be achieved with a moveNode (see 5.2.2) with a .null node as the direction (destination node).

#### 5.2.4.1 Request

- A HTTP DELETE to the URL of an existing node

#### 5.2.4.2 Response

- A HTTP 204 status code

#### 5.2.4.3 Faults

- The service SHOULD throw a HTTP 500 status code including an InternalFault fault in the entity-body if the operation fails
- The service SHALL throw a HTTP 401 status code including a PermissionDenied fault in the entity-body if the user does not have permissions to perform the operation
- The service SHALL throw a HTTP 404 status code including a NodeNotFound fault in the entity-body if the target node does not exist
- If a parent node in the URI path does not exist then the service SHALL throw a HTTP 404 status code including a ContainerNotFound fault in the entity-body
  - For example, given the URI path /a/b/c, the service must throw a HTTP 404 status code including a ContainerNotFound fault in the entity-body if either /a or /a/b do not exist.
- If a parent node in the URI path is a LinkNode, the service SHALL throw a HTTP 400 status code including a LinkFound fault in the entity-body.
  - For example, given the URI path /a/b/c, the service must throw a HTTP 400 status code including a LinkFound fault in the entity-body if either /a or /a/b are LinkNodes.
- If the target node in the URI path does not exist, the service SHALL throw a HTTP 404 status code including a NodeNotFound fault in the entity-body.
  - For example, given the URI path /a/b/c, the service must throw a HTTP 404 status code including a NodeNotFound fault in the entity-body if /a/b/c does not exist.

```
Example: deleteNode

  > curl -v -X DELETE "http://localhost:8000/nodes/mydata1"

* Connected to localhost (127.0.0.1) port 8000 (#0)
> DELETE /nodes/mydata1 HTTP/1.1
> User-Agent: curl/7.19.4 (universal-apple-darwin10.0) libcurl/7.19.4 OpenSSL/0.9.8l zlib/1.2.3
> Host: localhost:8000
> Accept: */*
>
< HTTP/1.1 200 OK
< Date: Thu, 11 Mar 2010 22:08:22 GMT
< Content-Length: 0
< Content-Type: text/html
< Allow: DELETE, GET, HEAD, POST, PUT
< Server: CherryPy/3.1.2
<
* Connection #0 to host localhost left intact
* Closing connection #0
```

## 5.3 Accessing metadata

### 5.3.1 getNode

Get the details for a specific Node.

#### 5.3.1.1 Request

- A HTTP GET to the URL of an existing node http://rest-endpoint/nodes/path

This can take the optional parameters to control the size of the response:

- *detail* with values of:
  - min: the returned record for the node contains minimum detail with all optional parts removed - the node type should be returned
    - e.g. <Node uri="vos://service/name" xsi:type="Node"/>
  - max: the returned record for the node contains the maximum detail, including any xsi:type specific extensions
  - properties: the returned record for the node contains the basic node element with a list of properties but no xsi:type specific extensions
- *uri* with a value of a VOSpace identifier, URI-encoded according to RFC2396
- *limit* with an integer value indicating the maximum number of results in the response.
  - No limit indicates a request for an unpaged list. However the server MAY still impose its own limit on the size of an individual response, splitting the results into more than one page if required/

The list of supported protocols for a node can be retrieved with:

- A HTTP POST of an empty protocol representation to the URL of an existing node http://rest-endpoint/nodes/path

#### 5.3.1.2 Response

- A HTTP 200 status code and a Node representation in the entity-body

When no parameters are present in the request, the full expanded record for the node SHALL be returned, including any xsi:type specific extensions; otherwise the appropriate form for the specified value of the "detail" parameter SHALL be returned.

When the node is a container, the returned record will also contain a list of direct children nodes in the container (as <node> subelements under the < nodes> element).

If a "uri" and "offset" are specified in the request then the returned list will consist of the subset of children which begins at the node matching the specified value of the "uri" parameter and with cardinality matching the specified value of the "offset" parameter drawn from an ordered sequence of all children. The ordering is determined by the server but results must always be drawn from the same ordered sequence.

#### 5.3.1.3 Faults

- The service SHOULD throw a HTTP 500 status code including an InternalFault fault in the entity-body if the operation fails
- The service SHALL throw a HTTP 401 status code including a PermissionDenied fault in the entity-body if the user does not have permissions to perform the operation
- The service SHALL throw a HTTP 404 status code including a NodeNotFound fault in the entity-body if the target Node does not exist

---

**Example: getNode**

```
  > curl -v "http://localhost:8000/nodes/mydir?detail=min&uri=vos://nvo.caltech!vospace/mydir/ngc3401"

* Connected to localhost (127.0.0.1) port 8000 (#0)
> GET /nodes/mydir?detail=min&uri=vos://nvo.caltech!vospace/mydir/ngc3401 HTTP/1.1
> User-Agent: curl/7.19.4 (universal-apple-darwin10.0) libcurl/7.19.4 OpenSSL/0.9.8l zlib/1.2.3
> Host: localhost:8000
> Accept: */*
>
< HTTP/1.1 200 OK
< Date: Fri, 12 Mar 2010 04:05:39 GMT
< Content-Length: 593
< Content-Type: text/html
< Allow: DELETE, GET, HEAD, POST, PUT
< Server: CherryPy/3.1.2
<
<node xmlns:xsi="http://www.w3.org/2001/XMLSchema-instance"
xsi:type="vos:ContainerNode" uri="vos://nvo.caltech!vospace/mydir"
xmlns="http://www.ivoa.net/xml/VOSpace/v2.0">
    <properties>
      <property uri="ivo://ivoa.net/vospace/core#description">My award winning images</property>
    </properties>
    <accepts>
      <view uri="ivo://ivoa.net/vospace/core#anyview"/>
    </accepts>
    <provides>
      <view uri="ivo://ivoa.net/vospace/core#defaultview"/>
    </provides>
    <capabilities/>
    <nodes>
      <node uri="vos://nvo.caltech!vospace/mydir/ngc4323"/>
      <node uri="vos://nvo.caltech!vospace/mydir/ngc5796"/>
      <node uri="vos://nvo.caltech!vospace/mydir/ngc6801"/>
    </nodes>
</node>
* Connection #0 to host localhost left intact
* Closing connection #0
```

### 5.3.2 setNode

Set the property values for a specific Node

#### 5.3.2.1 Request

- A HTTP POST of a Node representation to the URL of an existing node http://rest-endpoint/nodes/path including:
  - node: A Node containing the Node uri and a list of the Properties to set (as defined in section 3.1)

This will add or update the node properties including any xsi:type specific elements.

The operation is the union of existing values and new ones.

- An empty value sets the value to blank.
- To delete a Property, set the xsi:nil attribute to true

This method cannot be used to modify the Node type.

This method cannot be used to modify the accepts or provides list of Views for the Node.

This method cannot be used to create children of a container Node.

#### 5.3.2.2 Response

A HTTP 200 status code and a Node representation in the entity-body

The full expanded record for the node SHALL be returned, including any xsi:type specific extensions.

#### 5.3.2.3 Faults

- The service SHOULD throw a HTTP 500 status code including an InternalFault fault in the entity-body if the operation fails
- The service SHALL throw a HTTP 401 status code including a PermissionDenied fault in the entity-body if the request attempts to modify a read-only Property
- The service SHALL throw a HTTP 401 status code including a PermissionDenied fault in the entity-body if the user does not have permissions to perform the operation
- The service SHALL throw a HTTP 404 status code including a NodeNotFound fault in the entity-body if the target Node does not exist
- If a parent node in the URI path does not exist then the service SHALL throw a HTTP 404 status code including a ContainerNotFound fault in the entity-body
  - For example, given the URI path /a/b/c, the service must throw a HTTP 404 status code including a ContainerNotFound fault in the entity-body if either /a or /a/b do not exist.
- The service SHALL throw a HTTP 400 status code including an InvalidArgument fault in the entity-body if a specified property value is invalid

---

**Example: setNode**

The updated node (newNode.xml) is:

```
<node xmlns:xsi="http://www.w3.org/2001/XMLSchema-instance"
xsi:type="vos:UnstructuredDataNode"
uri="vos://nvo.caltech!vospace/mydata1"
xmlns="http://www.ivoa.net/xml/VOSpace/v2.0">
  <properties>
    <property uri="ivo://ivoa.net/vospace/core#date">2010-03-12</property>
  </properties>
  <accepts/>
  <provides/>
  <capabilities/>
</node>

> curl -v -X POST -d @newNode.xml "http://localhost:8000/nodes/mydata1"

* Connected to localhost (127.0.0.1) port 8000 (#0)
> POST /nodes/mydata1 HTTP/1.1
> User-Agent: curl/7.19.4 (universal-apple-darwin10.0) libcurl/7.19.4 OpenSSL/0.9.8l zlib/1.2.3
> Host: localhost:8000
> Accept: */*
> Content-Length: 309
> Content-Type: application/x-www-form-urlencoded
>
< HTTP/1.1 200 OK
< Date: Fri, 12 Mar 2010 18:49:25 GMT
< Content-Length: 445
< Content-Type: text/html
< Allow: DELETE, GET, HEAD, POST, PUT
< Server: CherryPy/3.1.2
<
<node xmlns:xsi="http://www.w3.org/2001/XMLSchema-instance"
xsi:type="vos:UnstructuredDataNode" uri="vos://nvo.caltech!vospace/mydata1"
xmlns="http://www.ivoa.net/xml/VOSpace/v2.0" busy="false">
  <properties>
    <property uri="ivo://ivoa.net/vospace/core#description">My important results</property>
    <property uri="ivo://ivoa.net/vospace/core#date">2010-03-12</property>
  </properties>
  <accepts>
    <view uri="ivo://ivoa.net/vospace/core#anyview"/>
```

```
    </accepts>
    <provides/>
    <capabilities/>
</node>
* Connection #0 to host localhost left intact
* Closing connection #0
```

### 5.3.3 findNodes

Find nodes whose properties match the specified values.

This operation is OPTIONAL.

#### 5.3.3.1 Request

- A HTTP POST of a Job representation of a Search (see section 3.7) to http://rest-endpoint/searches

A null value of the "matches" parameter implies a full listing of the space.

#### 5.3.3.2 Response

- The initial response is a HTTP 303 status code with the Location header keyword assigned to the created job.

The search results representation can be retrieved directly from the link reported in the Results List, available from the results endpoint - http://rest-endpoint/searches/(jobid)/results -, the standard UWS location under the results endpoint - http://rest-endpoint/searches/(jobid)/results/searchDetails - (which may well just be a redirect to the former link), or as part of the full Job representation from http://rest-endpoint/searches/(jobid). The result element in the Results List SHALL have an id of "searchDetails":

```
<uws:job>
...
   <uws:results>
     <uws:result id="searchDetails" xlink:href="http://rest-endpoint/searches/(jobid)/results/listing1"/>
   </uws:results>
...
</uws:job>
```

#### 5.3.3.3 Faults

For all faults, the service shall set the PHASE to "ERROR" in the Job representation. The <errorSummary> element in the Job representation shall be set to the appropriate value for the fault type and the appropriate fault representation (see section 5.5) provided at the error URI: http://rest-endpoint/transfers/(jobid)/error.

| Fault description | errorSummary | Fault representation |
|---|---|---|
| Operation not supported | Operation Not Supported | OperationNotSupported |
| Operation fails | Internal Fault | InternalFault |
| User does not have permissions to perform the operation | Permission Denied | PermissionDenied |
| A particular property is specified and does not exist in the space | Property Not Found | PropertyNotFound |

---

**Example: findNodes**

The Job to be submitted (newJob.xml) is:

```
<vos:search xmlns:vos="http://www.ivoa.net/xml/VOSpace/v2.0">
   <vos:detail>properties</vos:detail>
   <vos:matches>ivo://ivoa.net/vospace/core#description='galax'</vos:matches>
<vos:search>

> curl -v -X POST -d @newJob.xml "http://localhost:8000/searches"

* Connected to localhost (127.0.0.1) port 8000 (#0)
> POST /searches HTTP/1.1
> User-Agent: curl/7.19.4 (universal-apple-darwin10.0) libcurl/7.19.4 OpenSSL/0.9.8l zlib/1.2.3
> Host: localhost:8000
> Accept: */*
> Content-Length: 684
> Content-Type: application/x-www-form-urlencoded
>
< HTTP/1.1 303 See Other
< Content-Length: 172
< Server: CherryPy/3.1.2
< Location: http://localhost:8080/searches/8c5b0f78cd5a44af8694f10da1b92060
< Allow: DELETE, GET, HEAD, POST, PUT
< Date: Fri, 12 Mar 2010 19:50:12 GMT
< Content-Type: text/html
<
* Connection #0 to host localhost left intact
* Closing connection #0
This resource can be found at http://localhost:8080/searches/8c5b0f78cd5a44af8694f10da1b92060
```

The status of the job can now be polled at the job location:

```
> curl -v "http://localhost:8000/searches/8c6e7bc53ee848638cda35817e47da65"
```

```
* Connected to localhost (127.0.0.1) port 8000 (#0)
> GET /searches/8c6e7bc53ee848638cda35817e47da65 HTTP/1.1
> User-Agent: curl/7.19.4 (universal-apple-darwin10.0) libcurl/7.19.4 OpenSSL/0.9.8l zlib/1.2.3
> Host: localhost:8000
> Accept: */*
>
< HTTP/1.1 200 OK
< Date: Fri, 12 Mar 2010 19:51:24 GMT
< Content-Length: 891
< Content-Type: text/html
< Allow: DELETE, GET, HEAD, POST, PUT
< Server: CherryPy/3.1.2
<
<uws:job xmlns:uws="http://www.ivoa.net/xml/UWS/v1.0" xmlns:xlink="http://www.w3.org/1999/xlink"
  xmlns:xsi="http://www.w3.org/2001/XMLSchema-instance" xsi:schemaLocation="http://www.ivoa.net/xml/UWS/v1.0 UWS.xsd ">
  <uws:jobId>8c6e7bc53ee848638cda35817e47da65</uws:jobId>
  <uws:ownerId xsi:nil="true"/>
  <uws:phase>COMPLETED</uws:phase>
  <uws:startTime>2010-03-12T19:50:56.552278</uws:startTime>
  <uws:endTime>2010-03-12T19:50:56.562416</uws:endTime>
  <uws:executionDuration>0</uws:executionDuration>
  <uws:destruction xsi:nil="true"/>
  <uws:jobInfo>
    <vos:search>
      <vos:detail>properties</vos:detail>
      <vos:matches>ivo://ivoa.net/vospace/core#description='galax'</vos:matches>
    </vos:search>
  </uws:jobInfo>
  <uws:results>
    <uws:result id="searchDetails" xlink:href="http://localhost:8000/searches/d55814f88d974c21afe5ad50e4e875c8/results/listing1"/>
  </uws:results>
</uws:job>
* Connection #0 to host localhost left intact
* Closing connection #0
```

Once the Job has completed, the result can be obtained from the URL reported in the result element:

```
> curl -v "http://localhost:8000/searches/d55814f88d974c21afe5ad50e4e875c8/results/listing1"
```

```
* Connected to localhost (127.0.0.1) port 8000 (#0)
> GET /searches/d55814f88d974c21afe5ad50e4e875c8/results/listing1 HTTP/1.1
> User-Agent: curl/7.19.4 (universal-apple-darwin10.0) libcurl/7.19.4 OpenSSL/0.9.8l zlib/1.2.3
> Host: localhost:8000
> Accept: */*
>
< HTTP/1.1 200 OK
< Date: Fri, 12 Mar 2010 20:29:25 GMT
< Content-Length: 586
< Content-Type: text/html
< Allow: DELETE, GET, HEAD, POST, PUT
< Server: CherryPy/3.1.2
<
<nodes xmlns="http://www.ivoa.net/xml/VOSpace/v2.0"
xmlns:xsi="http://www.w3.org/2001/XMLSchema-instance">
  <node uri="vos://nvo.caltech!vospace/mydir/img1" xsi:type="vos:UnstructuredDataNode">
    <properties>
      <property uri="ivo://ivoa.net/vospace/core#description">This is an R-band image of the galaxy NGC 3276</property>
    </properties>
  </node>
  <node uri="vos://nvo.caltech!vospace/mydir/img5" xsi:type="vos:StructuredDataNode">
    <properties>
      <property uri="ivo://ivoa.net/vospace/core#description">This is a Chandra mosaic of the Fornax cluster of galaxies</property>
    </properties>
  </node>
</nodes>
* Connection #0 to host localhost left intact
* Closing connection #0
```

## 5.4 Transferring data

Two modes are supported for external data transfers: a simple HTTP GET to retrieve data from a service (pullFromVoSpace) and a more general mechanism which employs a UWS-based approach [UWS] for submitting general data transfer requests (see section 3.6). In the latter, four directions are specified in which external data transfers can happen:

- sending data to a service (pushToVoSpace)
- importing data into a service (pullToVoSpace)
- reading data from a service (pullFromVoSpace)
- sending data from a service (pushFromVoSpace)

A transfer job is created by a HTTP POST of an appropriate Job representation to the transfers endpoint of the service: http://rest-endpoint/transfers

The service returns the jobid of the transfer and it can then be initiated with a HTTP POST of the single parameter "PHASE=RUN" to the appropriate job endpoint: http://rest-endpoint/transfers/(jobid)/phase. Alternatively the transfer can also be run immediately on creation by adding a "PHASE=RUN" to the initial Job representation.

The status of any transfer can be obtained by polling the phase endpoint for a particular transfer, i.e. a HTTP GET to http://rest-endpoint/transfers/(jobid)/phase.

Once a transfer has completed, any results can be obtained by following the link in the Results List available from the results endpoint for that transfer, i.e. with a HTTP GET to http://rest-endpoint/transfers/(jobid)/results. This pertains particularly to the transfer methods in which data is sent to or read from a service-negotiated URL (pushToVoSpace and pullFromVoSpace).

A transfer can also be aborted at any stage with a HTTP POST of the parameter "PHASE=ABORT" to the endpoint: http://rest-endpoint/transfers/(jobid)/phase

More specific details for each of the four directions are given below.

### 5.4.1 pushToVoSpace

Request a list of URLs to send data to a VOSpace node.

This method asks the server for a list of one or more URLs that the client can use to send data to.

The data transfer is initiated by the client, after it has received the response from the VOSpace service.

The primary use case for this method is client that wants to send some data directly to a VOSpace service.

This operation is OPTIONAL.

#### 5.4.1.1 Request

A HTTP POST of a Job representation for the transfer to http://rest-endpoint/transfers

If a Node already exists at the target URI, then the data SHALL be imported into the existing Node and the Node properties SHALL be cleared unless the node is a ContainerNode.

If there is no Node at the target URI, then the service SHALL create a new Node using the uri and the default xsi:type for the space.

The transfer representation contains details of the View and a list of the Protocols that the client wants to use.

The list of Protocols SHOULD not contain endpoint addresses, the service will supply the endpoint addresses in the response.

The service SHALL ignore any of the requested protocols that it does not understand or is unable to support.

.auto is the reserved URI to indicate an auto-generated URI for the destination, i.e. vos://service/path/.auto SHALL cause a new unique URI for the node within vos://service/path to be generated.

There is also an alternate convenience mode:

- A HTTP POST of a Job representation for the transfer to http://rest-endpoint/sync

The assumed transfer protocol is HTTP PUT; however, transfer negotiation in the usual manner is possible with this convenience mode.

#### 5.4.1.2 Response

- The initial response is a HTTP 303 status code with the Location header keyword assigned to the created job.

The service SHALL select which of the requested Protocols it is willing to provide and fill in the operational details for each one in the transfer result representation - this SHOULD normally include specifying the destination URL of the transfer protocol endpoint.

The transfer result SHOULD not include any Protocols that it is unable to support.

The transfer results representation can be retrieved directly from the link reported in the Results List, available either from the results endpoint - http://rest-endpoint/transfers/(jobid)/results - or as part of the full Job representation for the completed transfer available from http://rest-endpoint/transfers/(jobid). The result element in the Results List SHALL have an id of "transferDetails":

```
<uws:job>
...
  <uws:results>
    <uws:result id="transferDetails" xlink:href="http://rest-endpoint/transfers/(jobid)/results/details1"/>
  </uws:results>
...
</uws:job>
```

For the alternate convenience mode:

- A HTTP 303 status code with the Location header keyword assigned to the endpoint: http://rest-endpoint/transfers/(jobid)/results/transferDetails.

The HTTP 303 redirect points to a transfer representation with the required transfer details and endpoints.

#### 5.4.1.3 Faults

For all faults using the UWS mode, the service shall set the PHASE to "ERROR" in the Job representation. The <errorSummary> element in the Job representation shall be set to the appropriate value for the fault type and the appropriate fault representation (see section 5.5) provided at the error URI: http://rest-endpoint/transfers/(jobid)/error.

| Fault description | errorSummary | Fault representation |
|---|---|---|
| Operation not supported | | |

| | Operation Not Supported | OperationNotSupported |
|---|---|---|
| Operation fails | Internal Fault | InternalFault |
| User does not have permissions to perform the operation | Permission Denied | PermissionDenied |
| Service does not support the requested View | View Not Supported | ViewNotSupported |
| Service supports none of the requested Protocols | Protocol Not Supported | ProtocolNotSupported |
| A View parameter is invalid | Invalid Argument | InvalidArgument |
| A Protocol parameter is invalid | Invalid Argument | InvalidArgument |

If an error occurs with the alternate convenience mode, the resulting transfers document SHOULD have no protocols. The client can then retrieve the Job representation for error information as with asynchronous transfers.

---

**Example: pushToVoSpace**

The Job to be submitted (newJob.xml) is:

```
<vos:transfer xmlns:vos="http://www.ivoa.net/xml/VOSpace/v2.0">
    <vos:target>vos://nvo.caltech!vospace/mydata1</vos:target>
    <vos:direction>pushToVoSpace</vos:direction>
    <vos:view>ivo://net.ivoa/vospace/core#fits</vos:view>
    <vos:protocol uri="ivo//net.ivoa/vospace/core#httpput"/>
</vos:transfer>
```

```
> curl -v -X POST -d @newJob.xml "http://localhost:8000/transfers"

* Connected to localhost (127.0.0.1) port 8000 (#0)
> POST /transfers HTTP/1.1
> User-Agent: curl/7.19.4 (universal-apple-darwin10.0) libcurl/7.19.4 OpenSSL/0.9.8l zlib/1.2.3
> Host: localhost:8000
> Accept: */*
> Content-Length: 836
> Content-Type: application/x-www-form-urlencoded
>
< HTTP/1.1 303 See Other
< Content-Length: 174
< Server: CherryPy/3.1.2
< Location: http://localhost:8000/transfers/fd5cf0cb1b6d4fbd84602982abf19ef1
< Allow: DELETE, GET, HEAD, POST, PUT
< Date: Fri, 12 Mar 2010 22:12:21 GMT
< Content-Type: text/html
<
* Connection #0 to host localhost left intact
* Closing connection #0
This resource can be found at http://localhost:8000/transfers/fd5cf0cb1b6d4fbd84602982abf19ef1.
```

The status of the job can now be polled at the job location:

```
> curl -v 'http://localhost:8000/transfers/fd5cf0cb1b6d4fbd84602982abf19ef1'

* Connected to localhost (127.0.0.1) port 8000 (#0)
> GET /transfers/fd5cf0cb1b6d4fbd84602982abf19ef1 HTTP/1.1
> User-Agent: curl/7.19.4 (universal-apple-darwin10.0) libcurl/7.19.4 OpenSSL/0.9.8l zlib/1.2.3
> Host: localhost:8000
> Accept: */*
>
< HTTP/1.1 200 OK
< Date: Fri, 12 Mar 2010 22:45:46 GMT
< Content-Length: 1037
< Content-Type: text/html
< Allow: DELETE, GET, HEAD, POST, PUT
< Server: CherryPy/3.1.2
<
<uws:job xmlns:uws="http://www.ivoa.net/xml/UWS/v1.0" xmlns:xlink="http://www.w3.org/1999/xlink"
    xmlns:xsi="http://www.w3.org/2001/XMLSchema-instance" xsi:schemaLocation="http://www.ivoa.net/xml/UWS/v1.0 UWS.xsd">
    <uws:jobId>fd5cf0cb1b6d4fbd84602982abf19ef1</uws:jobId>
    <uws:ownerId xsi:nil="true"/>
    <uws:phase>COMPLETED</uws:phase>
    <uws:startTime>2010-03-12T22:45:25.568694</uws:startTime>
    <uws:endTime>2010-03-12T22:45:25.568840</uws:endTime>
    <uws:executionDuration>0</uws:executionDuration>
    <uws:destruction xsi:nil="true"/>
    <uws:jobInfo>
      <vos:transfer>
        <vos:target>vos://nvo.caltech!vospace/mydata1</vos:target>
        <vos:direction>pushToVoSpace</vos:direction>
        <vos:view>ivo://net.ivoa/vospace/core#fits</vos:view>
        <vos:protocol uri="ivo//net.ivoa/vospace/core#httpput"/>
      </vos:transfer>
    </uws:jobInfo>
    <uws:results>
      <uws:result id="transferDetails"
xlink:href="http://localhost:8000/transfers/fd5cf0cb1b6d4fbd84602982abf19ef1/results/details"/>
    </uws:results>
</uws:job>
* Connection #0 to host localhost left intact
* Closing connection #0
```

The final negotiated details of the transfer can be retrieved from the URL reported in the result element:

```
> curl -v "http://localhost:8000/transfers/fd5cf0cb1b6d4fbd84602982abf19ef1/results/details"
```

```
* Connected to localhost (127.0.0.1) port 8000 (#0)
> GET /transfers/fd5cf0cb1b6d4fbd84602982abf19ef1/results/details HTTP/1.1
> User-Agent: curl/7.19.4 (universal-apple-darwin10.0) libcurl/7.19.4 OpenSSL/0.9.8l zlib/1.2.3
> Host: localhost:8000
> Accept: */*
>
< HTTP/1.1 200 OK
< Date: Fri, 12 Mar 2010 22:46:17 GMT
< Content-Length: 316
< Content-Type: text/html
< Allow: DELETE, GET, HEAD, POST, PUT
< Server: CherryPy/3.1.2
<
<transfer>
    <target>vos://nvo.caltech!vospace/mydata1</target>
    <direction>pushToVoSpace</direction>
    <view>ivo://ivoa.net/vospace/core#fits</view>
    <protocol uri="ivo://ivoa.net/vospace/core#httpput">
        <endpoint>http://localhost:8000/data/d55814f88d974c21afe5ad50e4e875c8</endpoint>
    </protocol>
</transfer>
* Connection #0 to host localhost left intact
* Closing connection #0
```

The FITS image can then be uploaded to the specified HTTP endpoint.

## 5.4.2 pullToVoSpace

Import data into a VOSpace node.

This method asks the server to fetch data from a remote location.

The data transfer is initiated by the VOSpace service and transferred direct into the target Node.

The data source can be another VOSpace service, or a standard HTTP or FTP server.

The primary use case for this method is transferring data from one server or service to another.

This operation is OPTIONAL.

### 5.4.2.1 Request

- A HTTP POST of a Job representation for the transfer to http://rest-endpoint/transfers

If a Node already exists at the target URI, then the data SHALL be imported into the existing Node and the Node properties SHALL be cleared unless the node is a ContainerNode.

If there is no Node at the target URI, then the service SHALL create a new Node using the uri, and the default xsi:type for the space.

### 5.4.2.2 Response

- The initial response is a HTTP 303 status code with the Location header keyword assigned to the created job.

### 5.4.2.3 Faults

For all faults, the service shall set the PHASE to "ERROR" in the Job representation. The <errorSummary> element in the Job representation shall be set to the appropriate value for the fault type and the appropriate fault representation (see section 5.5) provided at the error URI: http://rest-endpoint/transfers/(jobid)/error.

| Fault description | errorSummary | Fault representation |
|---|---|---|
| Operation not supported | Operation Not Supported | OperationNotSupported |
| Operation fails | Internal Fault | InternalFault |
| User does not have permissions to perform the operation | Permission Denied | PermissionDenied |
| Service does not support the requested View | View Not Supported | ViewNotSupported |
| Service supports none of the requested Protocols | Protocol Not Supported | ProtocolNotSupported |
| Destination URI is invalid | Invalid URI | InvalidURI |
| A View parameter is invalid | Invalid Argument | InvalidArgument |
| A Protocol parameter is invalid | Invalid Argument | InvalidArgument |
| Data format does not match the requested View | Invalid Data | InvalidData |

### 5.4.2.4 Notes

If the Job request contains more than one protocol parameter, then the service MAY fail over to use one or more of the options if the first one fails. The service SHOULD try each protocol option in turn until one succeeds or all have been tried.

**Example: pullToVoSpace**

The Job to be submitted (newJob.xml) is:

```
    <vos:transfer xmlns:vos="http://www.ivoa.net/xml/VOSpace/v2.0">
      <vos:target>vos://nvo.caltech!vospace/mydata1</vos:target>
      <vos:direction>pullToVoSpace</vos:direction>
      <vos:view>ivo://ivoa.net/vospace/core#fits</vos:view>
      <vos:protocol uri="ivo://ivoa.net/vospace/core#httpget">
        <vos:protocolEndpoint>http://some.server.com/here/is/the/data</vos:protocolEndpoint>
        </vos:protocol>
    </vos:transfer>

    > curl -v -X POST -d @newJob.xml "http://localhost:8000/transfers"

* Connected to localhost (127.0.0.1) port 8000 (#0)
> POST /transfers HTTP/1.1
> User-Agent: curl/7.19.4 (universal-apple-darwin10.0) libcurl/7.19.4 OpenSSL/0.9.8l zlib/1.2.3
> Host: localhost:8000
> Accept: */*
> Content-Length: 932
> Content-Type: application/x-www-form-urlencoded
>
< HTTP/1.1 303 See Other
< Content-Length: 174
< Server: CherryPy/3.1.2
< Location: http://localhost:8000/transfers/ea901be0f7ef41668df9916ca25820f8
< Allow: DELETE, GET, HEAD, POST, PUT
< Date: Fri, 12 Mar 2010 23:36:58 GMT
< Content-Type: text/html
<
* Connection #0 to host localhost left intact
* Closing connection #0
This resource can be found at http://localhost:8000/transfers/ea901be0f7ef41668df9916ca25820f8.
```

The status of the job can now be polled at the phase endpoint:

```
    > curl -v "http://localhost:8000/transfers/ea901be0f7ef41668df9916ca25820f8/phase"

* Connected to localhost (127.0.0.1) port 8000 (#0)
> GET /transfers/ea901be0f7ef41668df9916ca25820f8/phase HTTP/1.1
> User-Agent: curl/7.19.4 (universal-apple-darwin10.0) libcurl/7.19.4 OpenSSL/0.9.8l zlib/1.2.3
> Host: localhost:8000
> Accept: */*
>
< HTTP/1.1 200 OK
< Date: Fri, 12 Mar 2010 23:37:02 GMT
< Content-Length: 9
< Content-Type: text/html
< Allow: DELETE, GET, HEAD, POST, PUT
< Server: CherryPy/3.1.2
<
EXECUTING
* Connection #0 to host localhost left intact
* Closing connection #0
```

### 5.4.3 pullFromVoSpace

Request a set of URLs that the client can read data from.

The client requests access to the data in a Node, and the server SHALL respond with a set of URLs that the client can read the data from.

*5.4.3.1 Request*

- A HTTP POST of a Job representation for the transfer to http://rest-endpoint/transfers

The transfer representation SHOULD contain details of the View and a list of the Protocols that the client would like to use.

The list of Protocols SHOULD not contain endpoint addresses; the service SHALL supply the endpoint addresses in the response.

The service SHALL ignore any of the requested protocols that it does not understand or is unable to support.

A transfer may also be initiated by a HTTP POST of a Job representation for the transfer to http://rest-endpoint/sync. The assumed transport protocol is HTTP GET but transfer negotiation is possible.

There is also an alternate convenience mode (for example, for web browser access):

- A HTTP GET to the node URL that wants to be read with a parameter of "view = data":
  - http://rest-endpoint/nodes/path?view=data

The assumed transfer protocol is HTTP GET and the View of the node is its default value, i.e. no alternate Views are possible with this mode. Note that no transfer negotiation is possible with this convenience mode.

Other values of the URL parameter "view" may be used by individual services for particular purposes, e.g., "view=rss" to identify an RSS feed on the resource.

*5.4.3.2 Response*

- The initial response for the UWS mode is a HTTP 303 status code with the Location header keyword assigned to the created job

The service SHALL select which of the requested Protocols it is willing to provide and fill in the operational details for each one in the transfer result representation - this SHOULD normally include specifying the source URL of the transfer protocol endpoint.

The service response SHOULD not include any Protocols that it is unable to support.

The transfer results representation can be retrieved directly from the link reported in the Results List, available either from the results endpoint - http://rest-endpoint/transfers/(jobid)/results - or as part of the full Job representation for the completed transfer available from http://rest-endpoint/transfers/(jobid). In the latter case, the result element in the Results List SHALL have an id of "transferDetails":

```
<uws:job>
...
  <uws:results>
    <uws:result id="transferDetails" xlink:href="http://rest-endpoint/transfers/(jobid)/results/details1"/>
  </uws:results>
...
< /uws:job>
```

If the transfer was initiated with a HTTP POST to http://rest-endpoint/sync then the response is:

- A HTTP 303 status code with the Location header keyword assigned to the endpoint: http://rest-endpoint/transfers/(jobid)/results/transferDetails.

The HTTP 303 redirect points to a transfer representation with the required transfer details and endpoints.

For the alternate convenience mode (HTTP GET to the node):

- A direct byte stream for the node

or

- A HTTP 303 status code with the Location header keyword assigned to an endpoint where a direct byte stream can be obtained

Although HTTP GET is the assumed transfer protocol in this mode, the HTTP 303 redirect could point to a endpoint supported by an alternate transfer protocol, i.e. a FTP endpoint. The client can check for this on the scheme returned URI.

*5.4.3.3 Faults*

For all faults using the UWS mode, the service shall set the PHASE to "ERROR" in the Job representation. The <errorSummary> element in the Job representation shall be set to the appropriate value for the fault type and the appropriate fault representation (see section 5.5) provided at the error URI: http://rest-endpoint/transfers/(jobid)/error.

| Fault description | errorSummary | Fault representation |
|---|---|---|
| Operation fails | Internal Fault | InternalFault |
| User does not have permissions to perform the operation | Permission Denied | PermissionDenied |
| Source Node does not exist | Node Not Found | NodeNotFound |
| Service does not support the requested View | View Not Supported | ViewNotSupported |
| Service supports none of the requested Protocols | Protocol Not Supported | ProtocolNotSupported |
| A View parameter is invalid | Invalid Argument | InvalidArgument |
| A Protocol parameter is invalid | Invalid Argument | InvalidArgument |

If an error occurs with the alternate convenience mode, the resulting transfers document SHOULD have no protocols. The client can then retrieve the Job representation for error information as with asynchronous transfers.

*5.4.3.4 Notes*

The endpoint URLs supplied in the UWS response SHOULD be considered as 'one shot' URLs. A VOSpace service connected to a standard web server MAY return the public URL for the data. However, a different implementation MAY create a unique single use URL specifically for this transfer.

---

**Example: pullFromVoSpace**

The Job to be submitted (newJob.xml) is:

```
<vos:transfer xmlns:vos="http://www.ivoa.net/xml/VOSpace/v2.0">
  <vos:target>vos://nvo.caltech!vospace/mydata1</vos:target>
  <vos:direction>pullFromVoSpace</vos:direction>
  <vos:view ivo://ivoa.net/vospace/core#defaultview</vos:view>
  <vos:protocol uri="ivo://ivoa.net/vospace/core#httpget"/>
  <vos:protocol uri="ivo://ivoa.net/vospace/core#ftpget"/>
</vos:transfer>

> curl -v -X POST -d @newJob.xml "http://localhost:8000/transfers"

* Connected to localhost (127.0.0.1) port 8000 (#0)
> POST /transfers HTTP/1.1
> User-Agent: curl/7.19.4 (universal-apple-darwin10.0) libcurl/7.19.4 OpenSSL/0.9.8l zlib/1.2.3
> Host: localhost:8000
> Accept: */*
> Content-Length: 928
```

```
> Content-Type: application/x-www-form-urlencoded
>
< HTTP/1.1 303 See Other
< Content-Length: 174
< Server: CherryPy/3.1.2
< Location: http://localhost:8000/transfers/1ce91a4346b54432a4fd4a756fc75397
< Allow: DELETE, GET, HEAD, POST, PUT
< Date: Fri, 12 Mar 2010 23:46:46 GMT
< Content-Type: text/html
<
* Connection #0 to host localhost left intact
* Closing connection #0
This resource can be found at http://localhost:8000/transfers/1ce91a4346b54432a4fd4a756fc75397.
```

Once the job has completed (by polling the phase endpoint of the service), the results can be obtained from the result endpoint:

```
> curl -v "http://localhost:8000/transfers/1ce91a4346b54432a4fd4a756fc75397/results"

* Connected to localhost (127.0.0.1) port 8000 (#0)
> GET /transfers/1ce91a4346b54432a4fd4a756fc75397/results HTTP/1.1
> User-Agent: curl/7.19.4 (universal-apple-darwin10.0) libcurl/7.19.4 OpenSSL/0.9.8l zlib/1.2.3
> Host: localhost:8000
> Accept: */*
>
< HTTP/1.1 200 OK
< Date: Sat, 13 Mar 2010 00:00:20 GMT
< Content-Length: 239
< Content-Type: text/html
< Allow: DELETE, GET, HEAD, POST, PUT
< Server: CherryPy/3.1.2
<
<ns0:results xmlns:ns0="http://www.ivoa.net/xml/UWS/v1.0">
    <ns0:result xmlns:ns1="http://www.w3.org/1999/xlink" id="transferDetails"
    ns1:href="http://localhost:8000/transfers/83c19a500b1c48108d631f1aa020e8bb/results/details"/>
</ns0:results>
* Connection #0 to host localhost left intact
* Closing connection #0
```

The final negotiated details of the transfer can then be retrieved from the URL reported in the result element:

```
> curl -v "http://localhost:8000/transfers/83c19a500b1c48108d631f1aa020e8bb/results/details"

* Connected to localhost (127.0.0.1) port 8000 (#0)
> GET /transfers/1ce91a4346b54432a4fd4a756fc75397/results/details HTTP/1.1
> User-Agent: curl/7.19.4 (universal-apple-darwin10.0) libcurl/7.19.4 OpenSSL/0.9.8l zlib/1.2.3
> Host: localhost:8000
> Accept: */*
>
< HTTP/1.1 200 OK
< Date: Sat, 13 Mar 2010 00:13:18 GMT
< Content-Length: 477
< Content-Type: text/html
< Allow: DELETE, GET, HEAD, POST, PUT
< Server: CherryPy/3.1.2
<
<transfer>
    <target>vos://nvo.caltech!vospace/mydata1</target>
    <direction>pullFromVoSpace</direction>
    <view>ivo://ivoa.net/vospace/core#defaultview</view>
    <protocol uri="ivo://ivoa.net/vospace/core#httpget">
        <endpoint>http://localhost:8000/data/d27282305c6746889691e914abab9403</endpoint>
    </protocol>
    <protocol uri="ivo://ivoa.net/vospace/core#ftpget">
        <endpoint>ftp://localhost:8000/data/0d4824049dd444a290f6a524323dbcd0</endpoint>
    </protocol>
</transfer>
* Connection #0 to host localhost left intact
* Closing connection #0
```

### 5.4.4 pushFromVoSpace

Ask the server to send data to a remote location.

The client supplies a list of URLs and asks the server to send the data to the remote location.

The transfer is initiated by the server, and the data transferred direct from the server to the remote location.

This operation is OPTIONAL.

*5.4.4.1 Request*

- A HTTP POST of a Job representation for the transfer to http://rest-endpoint/transfers

*5.4.4.2 Response*

- The initial response is a HTTP 303 status code with the Location header keyword assigned to the created job.

| Fault description | errorSummary | Fault representation |
|---|---|---|
| Operation not supported | Operation Not Supported | OperationNotSupported |
| Operation fails | Internal Fault | InternalFault |
| User does not have permissions to perform the operation | Permission Denied | PermissionDenied |
| Source Node does not exist | Node Not Found | NodeNotFound |
| Service does not support the requested View | View Not Supported | ViewNotSupported |
| Service supports none of the requested Protocols | Protocol Not Supported | ProtocolNotSupported |
| Destination URI is invalid | Invalid URI | InvalidURI |
| A Protocol parameter is invalid | Invalid Argument | InvalidArgument |
| Data transfer does not complete | Transfer Failed | TransferFailed |

*5.4.4.4 Notes*

If the Job request contains more than one protocol parameter then the service MAY fail over to use one or more of the options if the first one fails. The service SHOULD try each protocol option in turn until one succeeds or all have been tried.

---

**Example: pushFromVoSpace**

The Job to be submitted (newJob.xml) is:

```
<vos:transfer xmlns:vos="http://www.ivoa.net/xml/VOSpace/v2.0">
    <vos:target>vos://nvo.caltech!vospace/mydata1</uws:parameter>
    <vos:direction>pushFromVoSpace</uws:parameter>
    <vos:view>ivo://ivoa.net/vospace/core#defaultview</uws:parameter>
    <vos:protocol uri="ivo://ivoa.net/vospace/core#httpput">
        <vos:protocolEndpoint>http://some.server.com/put/the/data/here</vos:protocolEndpoint>
    </vos:protocol>
    <vos:protocol uri="ivo://ivoa.net/vospace/core#ftpput">
        <vos:protocolEndpoint>ftp://some.other.server.com/put/the/data/here</vos:protocolEndpoint>
    </vos:protocol>
</vos:transfer>
```

> **curl -v -X POST -d @newJob.xml "http://localhost:8000/transfers"**

```
* Connected to localhost (127.0.0.1) port 8000 (#0)
> POST /transfers HTTP/1.1
> User-Agent: curl/7.19.4 (universal-apple-darwin10.0) libcurl/7.19.4 OpenSSL/0.9.8l zlib/1.2.3
> Host: localhost:8000
> Accept: */*
> Content-Length: 942
> Content-Type: application/x-www-form-urlencoded
>
< HTTP/1.1 303 See Other
< Content-Length: 174
< Server: CherryPy/3.1.2
< Location: http://localhost:8000/transfers/346d5cd27a0d405a8311819c90818cbc
< Allow: DELETE, GET, HEAD, POST, PUT
< Date: Sat, 13 Mar 2010 00:41:20 GMT
< Content-Type: text/html
<
* Connection #0 to host localhost left intact
* Closing connection #0
This resource can be found at http://localhost:8000/transfers/346d5cd27a0d405a8311819c90818cbc.
```

The Job can then be started with:

> **curl -v -d PHASE=RUN http://localhost:8000/transfers/346d5cd27a0d405a8311819c90818cbc**

```
* Connected to localhost (127.0.0.1) port 8000 (#0)
> POST /transfers HTTP/1.1
> User-Agent: curl/7.19.4 (universal-apple-darwin10.0) libcurl/7.19.4 OpenSSL/0.9.8l zlib/1.2.3
> Host: localhost:8000
> Accept: */*
> Content-Length: 9
> Content-Type: application/x-www-form-urlencoded
>
```

The status of the job can now be polled at the phase endpoint:

> **curl -v "http://localhost:8000/transfers/346d5cd27a0d405a8311819c90818cbc/phase"**

```
* Connected to localhost (127.0.0.1) port 8000 (#0)
> GET /transfers/346d5cd27a0d405a8311819c90818cbc/phase HTTP/1.1
> User-Agent: curl/7.19.4 (universal-apple-darwin10.0) libcurl/7.19.4 OpenSSL/0.9.8l zlib/1.2.3
> Host: localhost:8000
> Accept: */*
>
< HTTP/1.1 200 OK
< Date: Fri, 12 Mar 2010 23:37:02 GMT
< Content-Length: 9
< Content-Type: text/html
< Allow: DELETE, GET, HEAD, POST, PUT
< Server: CherryPy/3.1.2
```

```
<
COMPLETED
* Connection #0 to host localhost left intact
* Closing connection #0
```

## 5.5 Fault arguments

Faults reported by a VOSpace service SHALL contain the following information:

**5.5.1 InternalFault**

This is thrown with a description of the cause of the fault.

**5.5.2 PermissionDenied**

This is thrown with a description of why the credentials (if any were provided) were rejected.

**5.5.3 InvalidURI**

This is thrown with details of the invalid URI.

**5.5.4 NodeNotFound**

This is thrown with the URI of the missing Node.

**5.5.5 DuplicateNode**

This is thrown with the URI of the duplicate Node.

**5.5.6 InvalidToken**

This is thrown with the invalid token.

**5.5.7 InvalidArgument**

This is thrown with a description of the invalid argument, including the View or Protocol URI and the name and value of the parameter that caused the fault.

**5.5.8 TypeNotSupported**

This is thrown with the QName of the unsupported type.

**5.5.9 ViewNotSupported**

This is thrown with the uri of the View.

**5.5.10 InvalidData**

This is thrown with any error message that the data parser produced.

**5.5.11 LinkFoundFault**

The fault details must contain the full details of the LinkNode.

## 6 Changes since last version

**From version 2.00-20120824:**

- Use of Resource Idenfifier and CDP in Architecture diagram clarified.
- StandardID defined for this version.
- Various formatting issues.
- Use of HTTP 500 status code clarified.
- HTTP response codes made consistent with RFC 2616.

**From version 2.00-20111202:**

- Clarified use of URI fragments
- Added appendix of standard properties
- Clarified multi-value properties
- Added 404 error codes to getNode and setNode for no parent container ensuring consistent behaviour
- Adjusted pushToVoSpace and pullFromVoSpace convenience methods for self consistency and closer matching to UWS
- Fault formatting.

**From version 2.00-20110628:**

- Corrected REST binding for synchronous transfer details
- Added endpoint for synchronous job POSTs
- Added properties mtime, ctime, btime
- Changed text to describe use of StandardsRegExt for standard properties, views, capabilities, tranfer protocols
- Changed error code on ContainerNotFound and LinkFound faults
- Added correct failure response for unsupported operations

**From version 2.00-20101112:**

- Examples amended with correct namespaces and root elements
- "~" listed as a valid URI separator character along with "!"
- Core property list added to
- Properties with multiple values clarified
- Optional operations clarified and indicated in conformance matrix
- getNode and findNode operation arguments made consistent
- Synchronous HTTP PUT method added
- createNode response changed to HTTP 200 from HTTP 201
- /{transferDetails} endpoint added
- Support for .auto made optional

**From version 2.00-20100323:**

- Updated UWS details: use uws:jobInfo instead of uws:parameters
- Added IVOA architecture text
- Removed text about WSDL/WADL

# Appendix A: Machine readable definitions

## A.1 Message schema

---

**XML Schema**

The requests and responses of a VOSpace 2.0 service shall adhere to the following XML Schema:

```
<?xml version="1.0" encoding="UTF-8"?>
<xs:schema targetNamespace="http://www.ivoa.net/xml/VOSpace/v2.0" elementFormDefault="qualified"
    attributeFormDefault="unqualified" xmlns:xs="http://www.w3.org/2001/XMLSchema"
    xmlns:vos="http://www.ivoa.net/xml/VOSpace/v2.0"
    xmlns:uws="http://www.ivoa.net/xml/UWS/v1.0"
    xmlns:xlink="http://www.w3.org/1999/xlink">
    <xs:import namespace="http://www.ivoa.net/xml/UWS/v1.0" schemaLocation="http://www.ivoa.net/xml/UWS/v1.0"/>

    <!-- ======== Node types ======== -->

    <xsd:complexType name="Node">
        <xsd:annotation>
            <xsd:documentation>
                The base class for all nodes.
            </xsd:documentation>
        </xsd:annotation>
        <xsd:sequence>
            <xsd:element name="properties" type="vos:PropertyList" minOccurs="0" maxOccurs="1">
                <xsd:annotation>
                    <xsd:documentation>
                        The list of node properties.
                    </xsd:documentation>
                </xsd:annotation>
            </xsd:element>
        </xsd:sequence>
        <xsd:attribute name="uri" type="xsd:anyURI" use="required">
            <xsd:annotation>
                <xsd:documentation>
                    The node identifier URI.
                </xsd:documentation>
            </xsd:annotation>
        </xsd:attribute>
    </xsd:complexType>

    <xsd:complexType name="DataNode">
        <xsd:annotation>
            <xsd:documentation>
                The base class for data nodes.
            </xsd:documentation>
        </xsd:annotation>
        <xsd:complexContent>
            <xsd:extension base="vos:Node">
                <xsd:sequence>
                    <xsd:element name="accepts" type="vos:ViewList" minOccurs="0" maxOccurs="1">
                        <xsd:annotation>
                            <xsd:documentation>
                                The list of views or data formats that this node can accept.
```

```xml
                    A simple unstructured node may accept data in any format.
                    A structured node may only accept data in specific formats.
                  </xsd:documentation>
                </xsd:annotation>
              </xsd:element>
              <xsd:element name="provides" type="vos:ViewList" minOccurs="0" maxOccurs="1">
                <xsd:annotation>
                  <xsd:documentation>
                    The list of views or data formats that this node can provide.
                    A simple unstructured node may only provide access to the data in the original format.
                    A structured node may provide different views of the data generated by the service.
                  </xsd:documentation>
                </xsd:annotation>
              </xsd:element>
              <xsd:element name="capabilities" type="vos:CapabilityList" minOccurs="0" maxOccurs="1">
                <xsd:annotation>
                  <xsd:documentation>
                    The list of capabilities that this node can support.
                  </xsd:documentation>
                </xsd:annotation>
              </xsd:element>
            </xsd:sequence>
            <xsd:attribute name="busy" type="xsd:boolean" use="optional" default="false">
              <xsd:annotation>
                <xsd:documentation>
                  A flag to indicate if the node content is available.
                  This will be set to false while the data is being imported,
                  or if the underlying service is busy.
                </xsd:documentation>
              </xsd:annotation>
            </xsd:attribute>
          </xsd:extension>
        </xsd:complexContent>
    </xsd:complexType>

    <xsd:complexType name="UnstructuredDataNode">
      <xsd:annotation>
        <xsd:documentation>
          An unstructured data node, containing unspecified content.
          The service does not need to understand or interpret the content.
          This type of node can accept any format, and only provides one view returning the original data.
        </xsd:documentation>
      </xsd:annotation>
      <xsd:complexContent>
        <xsd:extension base="vos:DataNode"/>
      </xsd:complexContent>
    </xsd:complexType>

    <xsd:complexType name="StructuredDataNode">
      <xsd:annotation>
        <xsd:documentation>
          A structured data node, containing a specific data format that the service has understands.
          This type of node may only accept specific data formats, and provide different views of the
          data generated by the service.
        </xsd:documentation>
      </xsd:annotation>
      <xsd:complexContent>
        <xsd:extension base="vos:DataNode"/>
      </xsd:complexContent>
    </xsd:complexType>

    <xsd:complexType name="ContainerNode">
      <xsd:annotation>
        <xsd:documentation>
          A container node containing any type of node.
        </xsd:documentation>
      </xsd:annotation>
      <xsd:complexContent>
        <xsd:extension base="vos:DataNode">
          <xsd:sequence>
            <xsd:element name="nodes" minOccurs="1" maxOccurs="1">
              <xsd:annotation>
                <xsd:documentation>
                  A list of the direct children that the container has.
                </xsd:documentation>
              </xsd:annotation>
              <xsd:complexType>
                <xsd:sequence>
                  <xsd:element name="node" type="vos:Node" minOccurs="0" maxOccurs="unbounded"/>
                </xsd:sequence>
              </xsd:complexType>
            </xsd:element>
          </xsd:sequence>
        </xsd:extension>
      </xsd:complexContent>
    </xsd:complexType>

    <xsd:complexType name="LinkNode">
      <xsd:annotation>
        <xsd:documentation>
          A node that points to another resource.
        </xsd:documentation>
      </xsd:annotation>
      <xsd:complexContent>
        <xsd:extension base="vos:Node">
```

```xml
            <xsd:sequence>
              <xsd:element name="target" type="xsd:anyURI" minOccurs="1" maxOccurs="1">
                <xsd:annotation>
                  <xsd:documentation>
                    The identifier for the object that the LinkNode points to.
                  </xsd:documentation>
                </xsd:annotation>
              </xsd:element>
            </xsd:sequence>
          </xsd:extension>
        </xsd:complexContent>
      </xsd:complexType>

      <!-- ======== Property types ======== -->

      <xsd:complexType name="Property">
        <xsd:simpleContent>
          <xsd:extension base="xsd:string">
            <xsd:attributeGroup ref="vos:PropertyAttributeGroup"/>
          </xsd:extension>
        </xsd:simpleContent>
      </xsd:complexType>

      <xsd:complexType name="PropertyList">
        <xsd:annotation>
          <xsd:documentation>
            A container element for a list of properties.
          </xsd:documentation>
        </xsd:annotation>
        <xsd:sequence>
          <xsd:element name="property" type="vos:Property" minOccurs="0" maxOccurs="unbounded" nillable="true"/>
        </xsd:sequence>
      </xsd:complexType>

      <xsd:complexType name="PropertyReference">
        <xsd:annotation>
          <xsd:documentation>
            A reference to a property description.
          </xsd:documentation>
        </xsd:annotation>
        <xsd:attributeGroup ref="vos:PropertyAttributeGroup"/>
      </xsd:complexType>

      <xsd:complexType name="PropertyReferenceList">
        <xsd:annotation>
          <xsd:documentation>
            A container element for a list of property references.
          </xsd:documentation>
        </xsd:annotation>
        <xsd:sequence>
          <xsd:element name="property" type="vos:PropertyReference" minOccurs="0" maxOccurs="unbounded" nillable="true"/>
        </xsd:sequence>
      </xsd:complexType>

      < xsd:attributeGroup name="PropertyAttributeGroup">
        < xsd:attribute name="uri" type="xsd:anyURI" use="required">
          < xsd:annotation>
            < xsd:documentation>
              If the property has been registered, then the URI should point to the registration document. Third party
              tools may use the urn:xxxx syntax to add unregistered properties.
            </xsd:documentation>
          </xsd:annotation>
        </xsd:attribute>
        < xsd:attribute name="readOnly" type="xsd:boolean" use="optional" default
="false">
          < xsd:annotation>
            < xsd:documentation>
              A flag to indicate if the property is considered read-only. Attempting to modify a read-only property
              should generate a PermissionDenied fault.
            </xsd:documentation>
          </xsd:annotation>
        </xsd:attribute>
        < xsd:attribute name="xsd:nill" type="xsd:boolean" use="optional" default
="false">
          < xsd:annotation>
            < xsd:documentation>
              A flag to indicate if the property is null and should be deleted.
            </xsd:documentation>
          </xsd:annotation>
        </xsd:attribute>
      </xsd:attributeGroup>

      <xsd:complexType name="GetPropertiesResponse">
        <xsd:sequence>
          <xsd:element name="accepts" type="vos:PropertyReferenceList">
            <xsd:annotation>
              <xsd:documentation>
                A list of identifiers for the properties that the service accepts and understands.
              </xsd:documentation>
            </xsd:annotation>
          </xsd:element>
          <xsd:element name="provides" type="vos:PropertyReferenceList">
            <xsd:annotation>
              <xsd:documentation>
                A list of identifiers for the properties that the service provides.
```

```
              </xsd:documentation>
            </xsd:annotation>
          </xsd:element>
          <xsd:element name="contains" type="vos:PropertyReferenceList">
            <xsd:annotation>
              <xsd:documentation>
                A list of identifiers for all the properties currently used by nodes within the service.
              </xsd:documentation>
            </xsd:annotation>
          </xsd:element>
        </xsd:sequence>
      </xsd:complexType>

      <!-- ======== View types ======== -->

      <xsd:complexType name="Param">
        <xsd:annotation>
          <xsd:documentation>
            A view or protocol parameter.
          </xsd:documentation>
        </xsd:annotation>
        <xsd:simpleContent>
          <xsd:extension base="xsd:string">
            <xsd:attribute name="uri" type="xsd:anyURI" use="required"/>
          </xsd:extension>
        </xsd:simpleContent>
      </xsd:complexType>

      <xsd:complexType name="View">
        <xsd:annotation>
          <xsd:documentation>
            An element describing a view of a data-set.
            A view may just provide the original data, or it could be server generated.
            Examples of server generated views could include a votable view of data in a database table,
            or a conversion from one image format to another.
          </xsd:documentation>
        </xsd:annotation>
        <xsd:sequence>
          <xsd:element name="param" type="vos:Param" minOccurs="0" maxOccurs="unbounded" nillable="true">
            <xsd:annotation>
              <xsd:documentation>
                A list of parameters for the view.
              </xsd:documentation>
            </xsd:annotation>
          </xsd:element>
        </xsd:sequence>
        <xsd:attribute name="uri" type="xsd:anyURI" use="required">
          <xsd:annotation>
            <xsd:documentation>
              The view URI.
              This should point to a resource describing the view format and what parameters it requires.
            </xsd:documentation>
          </xsd:annotation>
        </xsd:attribute>
        <xsd:attribute name="original" type="xsd:boolean" use="optional" default="true">
          <xsd:annotation>
            <xsd:documentation>
              A flag to indicate if the view provides access to the original data content or a derived form.
            </xsd:documentation>
          </xsd:annotation>
        </xsd:attribute>
      </xsd:complexType>

      <xsd:complexType name="ViewList">
        <xsd:annotation>
          <xsd:documentation>
            A container element for a list of views.
          </xsd:documentation>
        </xsd:annotation>
        <xsd:sequence>
          <xsd:element name="view" type="vos:View" minOccurs="0" maxOccurs="unbounded" nillable="true"/>
        </xsd:sequence>
      </xsd:complexType>

      <xsd:complexType name="GetViewsResponse">
        <xsd:sequence>
          <xsd:element name="accepts" type="vos:ViewList">
            <xsd:annotation>
              <xsd:documentation>
                A list of identifiers for the views that the service can accept.
                A simple file based system may accept data in 'any' format.
              </xsd:documentation>
            </xsd:annotation>
          </xsd:element>
          <xsd:element name="provides" type="vos:ViewList">
            <xsd:annotation>
              <xsd:documentation>
                A list of identifiers for the views that the service can provide.
                A simple file based system may only provide data in the original format.
              </xsd:documentation>
            </xsd:annotation>
          </xsd:element>
        </xsd:sequence>
      </xsd:complexType>
```

```xml
<!-- ======== Protocol types ======== -->

<xsd:complexType name="Protocol">
  <xsd:annotation>
    <xsd:documentation>
      A protocol element, containing the protocol URI, the endpoint and any protocol specific parameters.
    </xsd:documentation>
  </xsd:annotation>
  <xsd:sequence>
    <xsd:element name="endpoint" type="xsd:anyURI" minOccurs="0" maxOccurs="1">
      <xsd:annotation>
        <xsd:documentation>
          The target endpoint to use for a data transfer.
        </xsd:documentation>
      </xsd:annotation>
    </xsd:element>
    <xsd:element name="param" type="vos:Param" minOccurs="0" maxOccurs="unbounded" nillable="true">
      <xsd:annotation>
        <xsd:documentation>
          Any additional protocol specific parameters required to use the endpoint.
          For example, the user name or password to use for ftp access.
        </xsd:documentation>
      </xsd:annotation>
    </xsd:element>
  </xsd:sequence>
  <xsd:attribute name="uri" type="xsd:anyURI" use="required">
    <xsd:annotation>
      <xsd:documentation>
        The protocol identifier.
      </xsd:documentation>
    </xsd:annotation>
  </xsd:attribute>
</xsd:complexType>

<xsd:complexType name="ProtocolList">
  <xsd:annotation>
    <xsd:documentation>
      A container element for a list of protocols.
    </xsd:documentation>
  </xsd:annotation>
  <xsd:sequence>
    <xsd:element name="protocol" type="vos:Protocol" minOccurs="0" maxOccurs="unbounded" nillable="true"/>
  </xsd:sequence>
</xsd:complexType>

<xsd:complexType name="GetProtocolsResponse">
  <xsd:sequence>
    <xsd:element name="accepts" type="vos:ProtocolList">
      <xsd:annotation>
        <xsd:documentation>
          A list of identifiers for the protocols that the service can accept.
          This means that the service can act as a client for the protocol.
        </xsd:documentation>
      </xsd:annotation>
    </xsd:element>
    <xsd:element name="provides" type="vos:ProtocolList">
      <xsd:annotation>
        <xsd:documentation>
          A list of identifiers for the protocols that the service can provide.
          This means that the service can act as a server for the protocol.
        </xsd:documentation>
      </xsd:annotation>
    </xsd:element>
  </xsd:sequence>
</xsd:complexType>

<!-- ======== Capability types ======== -->

<xsd:complexType name="Capability">
  <xsd:annotation>
    <xsd:documentation>
      A capability element, containing the capability URI, the
      endpoint and any capability specific parameters(?).
    </xsd:documentation>
  </xsd:annotation>
  <xsd:sequence>
    <xsd:element name="endpoint" type="xsd:anyURI" minOccurs="0" maxOccurs="1">
      <xsd:annotation>
        <xsd:documentation>
          The target endpoint to use for the third-part interface.
        </xsd:documentation>
      </xsd:annotation>
    </xsd:element>
    <!--+
        | Uncommented Capability params.
        +-->
    <xsd:element name="param" type="vos:Param" minOccurs="0" maxOccurs="unbounded" nillable="true">
      <xsd:annotation>
        <xsd:documentation>
          Any additional capability specific parameters required to use the endpoint.
          For example, the user name or password to use for access.
        </xsd:documentation>
      </xsd:annotation>
    </xsd:element>
  </xsd:sequence>
```

```xml
        <xsd:attribute name="uri" type="xsd:anyURI" use="required">
          <xsd:annotation>
            <xsd:documentation>
              The capability identifier.
            </xsd:documentation>
          </xsd:annotation>
        </xsd:attribute>
      </xsd:complexType>

  <xsd:complexType name="CapabilityList">
      <xsd:annotation>
        <xsd:documentation>
          A container element for a list of capabilities.
        </xsd:documentation>
      </xsd:annotation>
      <xsd:sequence>
        <xsd:element name="capability" type="vos:Capability" minOccurs="0" maxOccurs="unbounded" nillable="true"/>
      </xsd:sequence>
  </xsd:complexType>

  <!-- ======== Node list type ======== -->

  <xsd:complexType name="NodeList">
     <xsd:annotation>
        <xsd:documentation>
          A container element for search responses.
        </xsd:documentation<
     </xsd:annotation>
     <xsd:sequence>
        <xsd:element name="nodes" minOccurs="0" maxOccurs="1">
           <xsd:annotation>
             <xsd:documentation>
               The list of nodes.
             </xsd:documentation>
           </xsd:annotation>
           <xsd:complexType>
              <xsd:sequence>
                <xsd:element name="node" type="vos:Node" minOccurs="0" maxOccurs="unbounded">
                  <xsd:annotation>
                    <xsd:documentation>
                      At the maximum level of detail this will be replaced by the full element for the extended type,
                      using xsi:type to indicate the node type/
                    </xsd:documentation>
                  </xsd:annotation>
                </xsd:element>
              </xsd:sequence>
           </xsd:complexType>
        </xsd:element>
     </xsd:sequence>
  </xsd:complexType>

  <!-- ======== Transfer type ======== -->

  <xs:complexType name="Transfer">
     <xsd:annotation>
        <xsd:documentation>
          A container element for transfer operations.
        </xsd:documentation<
     </xsd:annotation>
     <xs:sequence>
        <xs:element name="target" type="xs:anyURI" minOccurs="1" maxOccurs="1">
           <xsd:annotation>
             <xsd:documentation>
               The target of a transfer operation - the node to/from which data is to be transferred.
             </xsd:documentation>
           </xsd:annotation>
        </xs:element>
        <xsd:element name="direction" minOccurs="0" maxOccurs="1">
           <xsd:annotation>
             <xsd:documentation>
               The direction of a data transfer - either a URI or one of the specified directions
             </xsd:documentation>
           </xsd:annotation>
           <xsd:simpleType>
              <xsd:union>
                 <xsd:simpleType>
                    <xsd:restriction base="xsd:anyURI"/>
                 </xsd:simpleType>
                 <xsd:simpleType>
                    <xsd:restriction base="xs:string">
                       <xsd:enumeration value="pushToVoSpace"/>
                       <xsd:enumeration value="pushFromVoSpace"/>
                       <xsd:enumeration value="pullToVoSpace"/>
                       <xsd:enumeration value="pullFromVoSpace"/>
                    </xsd:restriction>
                 </xsd:simpleType>
              </xsd:union>
           </xsd:simpleType>
        </xsd:element>
        <xsd:element name="view" type="vos:View" minOccurs="0" maxOccurs="1">
           <xsd:annotation>
             <xsd:documentation>
               The requested view for the transfer.
             </xsd:documentation>
           </xsd:annotation>
```

```
        </xsd:element>
        <xsd:element name="protocol" type="vos:Protocol" minOccurs="0" maxOccurs="unbounded">
          <xsd:annotation>
            <xsd:documentation>
              The transfer protocol(s) to use.
            </xsd:documentation>
          </xsd:annotation>
        </xsd:element>
        <xsd:element name="keepBytes" type="xsd:boolean" minOccurs="0" maxOccurs="1">
          <xsd:annotation>
            <xsd:documentation>
              Indicates whether the source object is to be kept in an internal transfer, i.e., distinguishes between a move and a copy.
            </xsd:documentation>
          </xsd:annotation>
        </xsd:element>
      </xsd:sequence>
    </xsd:complexType>

    <!-- ======= Response representations ======== -->

    <xs:element name="protocols" type="vos:ProtocolList"/>

    <xs:element name="views" type="vos:ViewList"/>

    <xs:element name="properties" type="vos:PropertyList"/>

    <xs:element name="transfer" type="vos:Transfer"/>

    <xs:element name="searchDetails" type="vos:NodeList"/>
</xs:schema>
```

## Appendix B: Compliance matrix

This table summarizes the mandatory behaviour required of a fully compliant VOSpace 2.0 service, i.e. those operations denoted as SHALL and MUST occurring. Note that for faults the general condition is specified but specific details should be checked in the relevant sections. Those associated with optional features are marked with an asterisk in the appropriate section reference - if the service implements an optional feature then it must show this mandatory behaviour.

| Item | Description | Occurs in section(s) |
|---|---|---|
| 1 | VOSpace node identifiers are a URI with the scheme vos:// | 2 |
| 2 | The naming authority for a VOSpace node URI is the VOSpace service through which the node was created | 2 |
| 3 | "!" or "~" are used consistently in VOSpace node URIs | 2 |
| 4 | "!" or "~" are valid separator characters in service requests | 2 |
| 5 | All ancestors in a node hierarchy are resolvable as containers | 2 |
| 6 | The bit pattern for data stored in an *UnstructuredDataNode* is identical for read/write operations | 3.1[*] |
| 7 | A *Node* has elements: *uri, properties* and *capabilities* | 3.1 |
| 8 | A *DataNode* has elements: *uri, properties, capabilities, accepts, provides* and *busy* | 3.1 |
| 9 | A *LinkNode* has elements: *uri, properties, capabilities* and *target* | 3.1[*] |
| 10 | A *ContainerNode* and a *LinkNode* have no data bytes associated with them | 3.1[*] |
| 11 | A VOSpace service parses XML representations of all node types | 3.1 |
| 12 | A VOSpace services does not throw an XML parser error in response to requests about unsupported node types | 3.1 |
| 13 | A *Property* has elements:*uri, endpoint* and optional *readonly* flag | 3.2 |
| 14 | URIs must be valid (and unique) | 3.2.2, 3.3.2, 3.4.2, 3.5.2, 5.2.1.1, 5.4.1.1, 5.4.2.1 |
| 15 | A *Capability* has elements: *uri, endpoint* and *param* | 3.3 |
| 16 | Standard capabilities are represented by the specified URIs | 3.3.5 |
| 17 | Each *Property* is identified by a unique URI | 3.2.2 |
| 18 | Each *Capability* is identified by a unique URI | 3.3.2 |
| 19 | Each *View* is identified by a unique URI | 3.4.2 |
| 20 | Each *Protocol* is identified by a unique URI | 3.5.1 |
| 21 | Standard views are represented by the specified URIs | 3.4.4 |
| xx | Standard properties are represented by the specified URIs | 3.2.2 |
| 22 | Data imported with the default *View* is treated as a binary BLOB | 3.4.4.1 |
| 23 | Data exported with the default export View is returned in the most appropriate format | 3.4.4.2 |
| 24 | An archive format *View* on a *ContainerNode* provides access to the archive contents as children nodes of the container | 3.4.5[*] |
| 25 | An archive format *View* specified in a data export operation on a *ContainerNode* will package the contents of the container and all its child nodes | 3.4.5[*] |
| 26 | A *Protocol* has elements: *uri, endpoint<* and *param* | 3.5 |
| 27 | Standard protocols are represented by the specified URIs | 3.5.3 |

| 28 | A *Transfer* UWS Job representation has elements: *target, direction, view, protocol* and *keepBytes* | 3.6 |
|----|---|---|
| 29 | A *Transfer* results representation has elements: *target, direction, view,* and *protocol* | 3.6 |
| 30 | A server responds with a fault if it is unable to handle any of the requested protocols in a data transfer | 3.6.1, 3.6.2 |
| 31 | A server uses each requested protocol only once in a data transfer | 3.6.1, 3.6.2 |
| 32 | A data transfer is complete once a specified protocol is successful | 3.6.1, 3.6.2 |
| 33 | A data transfer has failed if none of the specified protocols has been successful | 3.6.1, 3.6.2 |
| 34 | A server updates the status flag in the UWS Job representation as appropriate | 3.6.1, 3.6.2 |
| 35 | A *Search* UWS Job representation has elements: *uri, limit, detail, matches* and *node* | 3.7[*] |
| 36 | A *Search* results representation has elements: *nodes* | 3.7[*] |
| 37 | A VOSpace service has the REST bindings: properties, views, protocols, searches, nodes, and transfers | 3.8 |
| 38 | Access policies on a VOSpace service are defined in the registered metadata for the service | 4 |
| 39 | VOSpace authentication employs IVOA SSO supported methods | 4 |
| 40 | A VOSpace service supports the operations: getProtocols, getViews, getProperties, createNode, deleteNode, moveNode, copyNode, getNode, setNode, pullFromVoSpace. The following are optional operations: findNodes, pushToVoSpace, pullToVoSpace, pushFromVoSpace | 5.1.1, 5.1.2, 5.1.3, 5.2.1, 5.2.2, 5.2.3, 5.2.4, 5.3.1, 5.3.2, 5.3.3[*], 5.4.1[*], 5.4.2[*], 5.4.3, 5.4.4[*] |
| 41 | *accepts* and *provides* specify entities that the service supports | 5.1.1.2, 5.1.2.2, 5.1.3.2 |
| 42 | A URI is autogenerated if the reserved URI `.auto` is used | 3.8[*], 5.2.1.1[*], 5.2.2.1[*], 5.2.3.1[*], 5.4.1.1[*] |
| 43 | An autogenerated URI is specified as a result in the Results List | 5.2.1.1[*], 5.2.2.1[*], 5.2.3.1[*], 5.4.1.1[*] |
| 44 | Any data written to the reserved URI `.null` is discarded | 3.8 |
| 45 | *accepts* is filled in based on service capabilities | 5.2.1.2 |
| 46 | If a container is deleted then so are its children | 5.2.4 |
| 47 | If a container is listed, only direct children are listed | 5.3.1.2 |
| 48 | Results are drawn from the same ordered sequence in any ordering imposed by the server | 5.3.1.2 |
| 49 | Regex expressions comply with POSIX conventions | 3.7 |
| 50 | Moving a container moves its children as well | 5.2.2 |
| 51 | Importing data into an existing container puts the new data within the container | 5.4.1 |
| 52 | Node types are not changed when a node is moved or copied | 5.2.2.2, 5.2.3.2 |
| 53 | Copying a container copies its children as well | 5.2.3.2 |
| 54 | Importing data into an existing node overwrites any preexisting data (unless it is a container) | 5.4.1.1[*], 5.4.2.1[*] |
| 55 | A HTTP 200 status code is returned if the operation completes successfully | 5.1.1.2, 5.1.2.2, 5.1.3.2, 5.2.1.2, 5.2.4, 5.3.1, 5.3.2 |
| 56 | A HTTP 303 status code is returned if the operation completes successfully | 5.2.2.2, 5.2.3.2, 5.3.3.2[*], 5.4.1.2[*], 5.4.2.2[*], 5.4.3.2, 5.4.4.2[*] |
| 57 | A HTTP 500 status code with an InternalFault fault in the body is thrown if the operation fails | 5.1.1.3, 5.1.2.3, 5.1.3.3, 5.2.1.3, 5.2.4.3, 5.3.1.3, 5.3.2.3, 5.4.3.3 |
| 58 | A HTTP 400 status code with an InvalidArgument fault in the body is thrown if a specified value is invalid | 5.3.2.3 |
| 59 | A HTTP 400 status code with an InvalidURI fault in the body is thrown if the specified URI is invalid | 5.2.1.3 |
| 60 | A HTTP 400 status code with a TypeNotSupported fault in the body is thrown if the specified type is not supported | 5.2.1.3 |
| 61 | A HTTP 401 status code with a PermissionDenied fault in the body is thrown if the user does not have permission to perform the operation | 5.2.1.3, 5.2.4.3, 5.3.1.3, 5.3.2.3, 5.4.3.3 |
| 62 | A HTTP 404 status code with a NodeNotFound fault in the body is thrown if... | 5.2.4.3, 5.3.1.3, 5.3.2.3, 5.4.3.3 |
| 63 | A HTTP 409 status code with a DuplicateFault fault in the body is thrown if the specified node already exists | 5.2.1.3 |
| 64 | A HTTP 404 status code with a ContainerNotFound fault in the body is thrown if a container is not found | 5.2.1.3, 5.2.4.3 |
| 65 | A HTTP 400 status code with a LinkFound fault in the body is thrown if a *LinkNode* is found | 5.2.1.3, 5.2.4.3 |
| 66 | The PHASE in the UWS Job representation is set to "ERROR" in case of a fault occurring | 5.2.2.3, 5.3.3.3[*], 5.4.1.3[*], 5.4.2.3[*], 5.4.3.3, 5.4.4.3[*] |
| 67 | The errorSummary element in the UWS Job representation is set to the appropriate value for the type of fault occurring | 5.2.2.3, 5.3.3.3[*], 5.4.1.3[*], 5.4.2.3[*], 5.4.3.3, 5.4.4.3[*] |
| 68 | The appropriate fault representation is provided at the error URI for the job | 5.2.2.3, 5.3.3.3[*], 5.4.1.3[*], 5.4.2.3[*], 5.4.3.3, 5.4.4.3[*] |

## Appendix C: Standard properties

This table summarizes the standard properties that SHALL be recognized by a VOSpace service and are registered in the .

| Property | Description | Data type |
|---|---|---|
| ivo://ivoa.net/vospace/core#availableSpace | the amount of space available within a container | string |
| ivo://ivoa.net/vospace/core#btime | the initial creation time | string |
| ivo://ivoa.net/vospace/core#contributor | an entity responsible for making contributions to this resource | URI |
| ivo://ivoa.net/vospace/core#coverage | a spatial or temporal topic of the resource, the spatial applicability of the resource, or the jurisdiction under which the resource is relevant | string |
| ivo://ivoa.net/vospace/core#creator | an entity primarily responsible for making the resource | URI |
| ivo://ivoa.net/vospace/core#ctime | the status change (aka metadata modification) time | datetime |
| ivo://ivoa.net/vospace/core#date | a point or period of time associated with an event in the lifecycle of the resource | datetime |
| ivo://ivoa.net/vospace/core#description | an account of the resource | string |
| ivo://ivoa.net/vospace/core#format | the file format, physical medium or dimensions of the resource | string |
| ivo://ivoa.net/vospace/core#groupread | the list of groups which can only read this resource | delimiter-separated list |
| ivo://ivoa.net/vospace/core#groupwrite | the list of groups which can read and write to this resource | delimiter-separated list |
| ivo://ivoa.net/vospace/core#identifier | an unambiguous reference to the resource within a given context | string/URI |
| ivo://ivoa.net/vospace/core#language | a language of the resource | string |
| ivo://ivoa.net/vospace/core#length | the length or size of a resource | float |
| ivo://ivoa.net/vospace/core#mtime | the data modification time | datetime |
| ivo://ivoa.net/vospace/core#publicread | whether this resource is world readable | boolean |
| ivo://ivoa.net/vospace/core#publisher | an entity responsible for making the resource available | uri |
| ivo://ivoa.net/vospace/core#quota | the value of a system quota on the resource | string |
| ivo://ivoa.net/vospace/core#relation | a related resource | uri |
| ivo://ivoa.net/vospace/core#rights | information about rights held in and over the resource | string |
| ivo://ivoa.net/vospace/core#source | a related resource from which the described resource is derived | string |
| ivo://ivoa.net/vospace/core#subject | the topic of the resource | string |
| ivo://ivoa.net/vospace/core#title | a name given to the resource | string |
| ivo://ivoa.net/vospace/core#type | the nature or genre of the resource | string |